**Review**

# From static structures to dynamic landscapes: cryo-EM redefines RNA biology

Shekhar Jadhav[1], Spandan Saha[1], Qingbin Shang[1], Marco Marcia[1,2,3,4,*]

[1]Department of Cell and Molecular Biology, Uppsala University, Husargatan 3, 75123 Uppsala, Sweden;

[2]Science for Life Laboratory, Department of Cell and Molecular Biology, Uppsala University, Husargatan 3, 75123 Uppsala, Sweden

[3]Istituto Italiano di Tecnologia, Via Morego 30, 16163, Genoa, Italy

[4]European Molecular Biology Laboratory (EMBL) Grenoble, 71 Avenue des Martyrs, Grenoble 38042, France

*Corresponding author:

Dr. Marco Marcia, Phone: +46184714697, E-mail: marco.marcia@icm.uu.se

## Abstract

RNA molecules perform diverse biological functions by dynamically exploring multiple conformational states rather than adopting a single static structure. Capturing these structural ensembles has long been a central challenge in molecular biology. Recent advances in cryogenic electron microscopy (cryo-EM) are now transforming this landscape by enabling the visualization of RNA molecules across a broad spectrum of functionally relevant conformations at near-atomic resolution.

Here, we examine how cryo-EM is reshaping RNA structural biology changing focus from the analysis of static structures to dynamic conformational landscapes. Through eight representative case studies spanning catalytic, regulatory, viral, and synthetic RNA assemblies, we illustrate how cryo-EM has revealed previously inaccessible mechanisms of RNA motion, including folding processes, ligand-dependent switching, and cooperative assembly. We specifically discuss emerging experimental and computational approaches that address and overcome the challenges associated with studying dynamic RNAs, particularly in construct design, sample preparation, vitrification, and data analysis. These novel methods resolve conformational variability and enable the reconstruction of discrete and continuous RNA conformational landscapes from cryo-EM data, highlighting how structural heterogeneity – once considered a limitation – can now be harnessed to extract functional insights.

Looking forward, the integration of cryo-EM with complementary biophysical techniques and time-resolved methodologies promises to bridge structural and temporal resolution, to routinely derive experimental molecular movies of RNA “in action”. These advances will not only deepen our understanding of RNA biology but also provide new opportunities for RNA-targeted therapeutics and the rational design of dynamic RNA-based nanodevices. By connecting structural snapshots into coherent dynamic models, cryo-EM is establishing a framework for quantitative descriptions of RNA energy landscapes and their functional roles.

## Introduction

Biological function emerges from the ability of well-defined molecular structures to explore dynamic conformational landscapes. Function arises from transitions between states, through coordinated motions that span multiple spatial and temporal scales (Henzler-Wildman & Kern, 2007). The concept of an energy landscape, in which macromolecules continuously interconvert between alternative conformations, provides a unifying framework to describe this dynamic behavior (Frauenfelder et al., 1991). But structural biology has so far primarily focused on capturing discrete conformational states of macromolecules.

RNA molecules exemplify this principle. Beyond adopting defined secondary and tertiary folds, many RNAs populate ensembles of conformations with comparable free energy, enabling them to act as molecular switches, scaffolds, catalysts, and regulators (Ganser et al., 2019; Vicens & Kieft, 2022). This intrinsic flexibility underlies key biological processes, from catalysis and gene regulation to viral replication and RNA-driven assembly. However, despite its fundamental importance, RNA dynamics has remained challenging to characterize at high resolution, due to both experimental and conceptual limitations.

Recent advances in biophysical techniques are now changing this landscape. Single-molecule fluorescence resonance energy transfer (smFRET) can detect hidden RNA states and enable increasely automatized RNA conformation screening, supporting RNA structure prediction (Groves et al., 2023; Hartmann et al., 2023; Weber et al., 2026). Molecular dynamics simulations (MD) and nuclear magnetic resonance (NMR) spectroscopy, especially when coupled to X-ray and neutron scattering (SAXS/SANS), can quantitatively describe the dynamics of low-populated RNA states essential for function (Arnold et al., 2025; Bussi et al., 2024). Single-molecule chemical probing further reveals the plasticity of RNA structure in response to cellular and environmental stimuli (Arney et al., 2024; Spitale & Incarnato, 2023). Integrating with and complementing these techniques, cryogenic electron microscopy (cryo-EM) is playing a transformative role, because it can now visualize discrete RNA conformational states and continuous structural variability at near atomic resolution (Khusainov et al., 2024). In other words, cryo-EM now visualizes in 3D RNA states that smFRET, SAXS/SANS and NMR can only infer kinetically and that chemical probing experiments can deconvolute but only at the level of secondary structure maps.

In contrast to traditional approaches that favor well-ordered, homogeneous samples, modern cryo-EM workflows – combined with innovations in sample design, vitrification, and computational analysis – are beginning to make the structural characterization of flexible and dynamic RNAs tractable (Bonilla & Jang, 2024; Bonilla & Kieft, 2022; Ma et al., 2022; Zhang et al., 2024). The opportunity to integrate cryo-EM now with biochemical, biophysical, computational and structural biology techniques marks a conceptual shift for RNA structural biology: from determining static structures toward reconstructing conformational ensembles and, ultimately, *energy landscapes*, intended as the multidimensional relationship between molecular conformations and free energy. This transition has been particularly rapid in recent years. The number of protein-free RNA structures determined by cryo-EM has increased sharply, driven by methodological advances that extend the technique to increasingly flexible and heterogeneous targets. At the same time, new computational tools enable the disentanglement of conformational heterogeneity and the *de novo* prediction of RNA 3D structures and conformational ensembles, revealing not only distinct structural states but also the motions that interconnect them. Together and particularly when complemented with computational (MD) and biophysical (NMR, SAXS/SANS, smFRET) studies, these cryo-EM approaches provide a direct structural view of RNA dynamics that was previously inaccessible, visualizing the distinct molecular configuration of energetically favorable states (valleys and minima of the *energy landscape* of the target) and enabling the calculation of the probability and timescale of conformational transitions (the barriers that separate the energy minima).

Here, we synthesize recent progress at this emerging interface between cryo-EM and RNA biology. We argue that cryo-EM is evolving into a central tool for decoding RNA structural-functional dynamics, enabling the visualization of both discrete states and continuous conformational trajectories. Through eight representative case studies spanning ribozymes, riboswitches, viral RNAs, and synthetic RNA assemblies, we illustrate how cryo-EM has revealed new mechanistic principles of RNA motion, including domain rearrangements, hinge-mediated flexibility, ligand-induced switching, and cooperative assembly. Building on these examples, we identify key experimental and computational bottlenecks – particularly in construct design, sample preparation, and data processing – and highlight recent solutions that have enabled the visualization of dynamic RNA behavior.

Finally, we outline future directions toward a more complete description of RNA dynamics, where cryo-EM data are integrated with complementary approaches such as molecular simulations, chemical probing, and time-resolved methods. These advances will not only deepen our understanding of RNA biology, but also open new opportunities for RNA-targeted therapeutics and the design of dynamic RNA-based nanodevices. By connecting structural snapshots into coherent dynamic models, cryo-EM is poised to move the field beyond static representations and toward a quantitative understanding of RNA energy landscapes.

## Structural studies on RNA

In 1871, Swiss chemist Friedrich Miescher described nuclein, later shown to contain both DNA and RNA, in his seminal work Über Die Chemische Zusammensetzung Der Eiterzellen (Miescher, 1871; Thess et al., 2021). Since then, it took over a century to discover messenger RNA (mRNA) (Astrachan & Volkin, 1958; Brenner et al., 1961; Gros et al., 1961; Hershey, 1953; Jacob & Monod, 1961; Monod et al., 1952; Pardee, 1954; Ycas & Vincent, 1960). Now, many RNA classes have been identified including tRNA, rRNA, microRNAs (miRNAs), small interfering RNAs (siRNAs), piwi-interacting RNA (piRNAs), small nuclear RNAs (snRNA), snoRNA (small nucleolar RNA), circular RNAs (circRNA) and long non-coding RNA (lncRNA) (Chen & Kim, 2024). Indeed, RNA comprises a major part of total cell biomass, up to 20% of a cell's dry weight, and the number of non-coding RNAs is greater than the number of proteins (Chen & Kim, 2024; Majumder et al., 2016). Importantly, the newly discovered classes of RNAs play fundamental roles in physiological gene regulation and in pathologies such as cancer, congenital syndromes and neurodegenerative disorders (Nemeth et al., 2024).

Despite the biological significance of RNA, structural studies of RNA have historically lagged behind structural studies on proteins. While the crystal structure of an RNA double helix was determined around the same time as that of the DNA double helix and of the first protein molecules (Rich & Watson, 1954), it took another 20 years to obtain the crystal structure of a full length, biologically relevant RNA, namely yeast phenylalanine tRNA ($tRNA^{Phe}$) (Robertus et al., 1974). But the functional importance of RNA structures started to be fully appreciated only after the discovery of catalytic RNAs in the '80s and the consequent determination of the crystal structure of the catalytic core of the *Tetrahymena* group I intron - which revealed an RNA-only active site (Altman et al., 1989; Cate et al., 1996; Kruger et al., 1982). In the late '90s, X-ray crystal structures of the 30S and 50S ribosomal subunits emphasized the potential of RNA to form a structural scaffold for proteins. At the same time, crystal structures of riboswitches revealed the ability of RNA to recognize small molecules selectively and specifically (Kavita & Breaker, 2023).

While the number of RNA structures is still now only representing less than 1 % of the number protein structures deposited in the PDB (as of Feb 5th 2026), recent advances in cryo-EM have enabled structure determination of monomeric, multimeric, synthetic, and chimeric RNAs. Since 2024, 126 new structures of RNAs have been determined by cryo-EM, more than doubling the number of available RNA cryo-EM structures. These studies have led to fundamental insights into the reaction mechanism of various ribozymes, such as self-splicing group I and group II introns, into gene regulation mechanisms of various riboswitches, into the potential of synthetic RNA-based nanotechnology, and into the mode of function of regulatory viral RNAs (vRNAs). Below, we briefly review the different classes of RNAs for which cryo-EM structures have been determined so far, and for each class we highlight a collection of significant recent case studies that specifically characterized RNA structure-functional dynamics.

### *Ribozymes*

RNA enzymes catalyze two of the most fundamental reactions in biology: protein synthesis and pre-mRNA splicing. These reactions are catalyzed by RNA in the context of large ribonucleoprotein (RNP) assemblies – the ribosome and the spliceosome.

Complementing decades of biochemical, single-molecule, and spectroscopic studies, recent advances in cryo-electron microscopy have provided unprecedented insight into the dynamic

| Box 1 - Ribozymes |
| --- |
| **Case study 1. Group I self-splicing introns**<br>*a. Mechanism of catalysis*<br>Key observations:<br>• in the first step of splicing, P1 helix (5'-exon paired to intron 5'-terminus) adopts three states (relaxed, intermediate and docked) (**Figure 1**)<br>• P1 moves by 66º from relaxed to intermediate state, and then by 50º from intermediate to docked state *via* interactions with J4/5 and J7/8 (**Figure 1**)<br>• P1 movement brings distance between 5'-exon and G cofactor from 62 to 4 Å (**Figure 1**)<br>• after 5'-exon cleavage, 3'-exon and intron 5'-end form P10 helix and 5'-exon, intron 5'-end and 3'-exon form a previously unknown pseudoknot.<br>Organism: *Tetrahymena thermophila*<br>Number of structures: 18<br>PDB id: 7YG9, 7YGA, 7YGB, 7YGC, 7YG8, 7YGD, 7XD3, 7XD4, 7XD5, 7XD6, 7XD7, 8HD6, 8HD7, 8I7N, 7YC8, 7YCG, 7YCH, 7YCI<br>Resolution: 2.7-4.1 Å<br>References: (Li et al., 2023; Luo et al., 2023; Zhang et al., 2023)<br><br>*b. Mechanism of folding*<br>Key observations:<br>• in the transition from the misfolded (M) intermediate to the folded, native (N) state:<br>  o P7 helix rotates by 90º and moves by 18.9 Å to dock on to the P9.1/P9.1a internal loop,<br>  o P3 helix completely refolds<br>  o P8 helix rotates by 360º around its helical axis to twist and correctly position J7/3 and J8/7<br>Organism: *Tetrahymena thermophila*<br>Number of structures: 5<br>PDB id: 7UVT, 7XSK, 7XSL, 7XSM, and 7XSN<br>Resolution: 3.5-4.0 Å<br>References: (Bonilla et al., 2022; S. Li et al., 2022) |
| **Case study 2. Group II self-splicing introns**<br>*a. Mechanism of catalysis*<br>Key observations:<br>• active site dynamics occurring during between first and second step of splicing, so far only hypothesized from crystal structures and mutagenesis experiments<br>Organism: *Oceanobacillus iheyensis*<br>Number of structures: 9<br>PDB id: 9QTK, 9QTV, 9QTW, 9QTX, 9QTZ, 9QU0, 9QU2, 9QU6 and 9QU7<br>Resolution: 2.6 Å<br>References: (Maiorca et al., 2026)<br><br>*b. Mechanism of folding*<br>Key observations:<br>• the intron 5'-terminal domain (domain 1, D1) folds first<br>• to dock the 3'-terminal domains (D2-D6), helical subdomain D1c rotates around conserved hinge residues displacing the tip of D1c by 51 Å<br>• D1 gating follows a "nutcracker" mechanism<br>Organism: *Oceanobacillus iheyensis*<br>Number of structures: 6<br>PDB id: 9G4I, 9G4J, 9G4L, 9G4V, 9G54, 9G56<br>Resolution: 2.6-7.6 Å<br>References: (Jadhav et al., 2025) |

behavior of these complexes, revealing conformational transitions that underlie substrate recognition, catalysis, and product release. For the ribosome, time-resolved cryo-EM and heterogeneous reconstructions – also *in situ* – have begun to capture transient functional states along the translation cycle (Garg et al., 2025; Rickgauer et al., 2024) and during early assembly (Qin et al., 2023). Moreover, cryo-EM has recently complemented classical crystallographic and NMR studies by revealing conformational heterogeneity in tRNA-containing complexes (Biela et al., 2023). Structural analyses have demonstrated that tRNAs undergo bending and twisting during ribosomal translocation and can adopt remodeled conformations upon interaction with processing and modification enzymes, underscoring the functional importance of their structural plasticity (Carbone et al., 2021; Loveland et al., 2020). Similarly, cryo-EM studies of the spliceosome coupled to computational simulations have uncovered a series of compositional and conformational rearrangements that accompany assembly and catalysis (Beusch & Madhani, 2024; Pokorna et al., 2025).

But recent work has begun to resolve at near-atomic resolution also the folding pathways and catalytic mechanisms of other splicing machines, namely self-splicing introns, which function as RNA-only systems. Self-splicing introns are ribozymes present in certain pre-mRNA transcripts of bacteria, archaea, and plants and fungal organelles. They catalyze the same two successive transesterification reactions as the spliceosome, to first cleave the 5'-exon and then ligate the 5' and 3' exons to generate a mature mRNA (Cech, 1990; Michel & Ferat, 1995). Among self-splicing introns, group I introns use a guanosine cofactor and form a linear intron product after excision (**Box 1 – Case study 1**, and **Figure 1**), whereas group II introns – which are evolutionary ancestors of the spliceosome – do not need cofactors and form a lariat product, i.e. a structure containing 2'-5' phosphodiester linkage at the so-called branchpoint adenosine (Cech, 1990; Xu et al., 2023) (**Box 1 – Case study 2**, and **Figure 2**). To perform splicing, both intron classes form intricate tertiary structures, wherein the active site coordinates mono- and divalent ions essential for splicing (Cate et al., 1996; Marcia & Pyle, 2012; Toor et al., 2008). For the last 40 years, both intron classes have been used as prototypical model systems for structural and biochemical studies on RNA.

### *Regulatory RNAs*

Beyond catalyzing fundamental biochemical reactions, RNA also plays key roles in the regulation of gene expression at both the transcriptional and translational levels.

Among RNA-only regulatory systems, riboswitches remain the best characterized. These non-coding and non-catalytic RNA elements, typically located in the 5′ untranslated regions of mRNAs, regulate gene expression by undergoing ligand-induced conformational changes that control transcription or translation (Garst et al., 2011). Using cryo-EM, various ligand-bound, ligand-free, active and inactive conformations of different riboswitches have been determined

| BOX 2 – REGULATORY RNAS |
|---|
| **Case study 3. Cobalamin riboswitch**<br>Key observations:<br>• P-, candy-, compact- and Y-shaped monomeric forms are formed in the absence of cobalamin; dimeric forms are formed in the presence of cobalamin through intermolecular kissing loops between P5 and P13 (**Figure 3**)<br>• increase in intramolecular P5-P3 distance from 11.6 Å (apo state) to 26.8 Å (dimer 1, holo state) and flexibility of P6 accompany ligand binding (**Figure 3**)<br>Organism: *Caldanaerobacter subterraneus subsp. tengcongensis*<br>Number of structures: 5<br>PDB id: 8SA2, 8SA3, 8SA4, 8SA5 and 8SA6<br>Resolution: 2.9-5.3 Å<br>References: (Ding, Deme, et al., 2023; Ding, Lee, et al., 2023) |

so far, including the SAM-IV riboswitch bound to S-adenosyl-methionine (SAM) (Zhang et al., 2019), the glycine riboswitch (Kappel et al., 2020), the T-box riboswitches (Jia et al., 2023; Li et al., 2019), and the cobalamin riboswitch (Ding, Deme, et al., 2023) (**Box 2 – Case study 3**, and **Figure 3**). Interestingly, because riboswitches are typically short and thus challenging for cryo-EM investigation, specific strategies have been put in place to overcome this limitation. For instance, the 35-kDa flavin mononucleotide (FMN) riboswitch was induced to oligomerize (Liu et al., 2022), whereas the fluoride riboswitch (Langeberg & Kieft, 2023) and the thiamine pyrophosphate (TPP) riboswitch (Haack et al., 2025) were fused to molecular scaffolds.

At the interface of RNA regulation and protein complexes, the 7SK ribonucleoprotein (7SK RNP) controls transcriptional elongation by modulating P-TEFb activity, and recent cryo-EM studies have begun to elucidate its structural organization and conformational variability, highlighting the role of RNA-mediated scaffolding in dynamic regulatory assemblies (Yang et al., 2022).

Furthermore, a prominent, recently discovered class of regulatory transcripts consists of long non-coding RNAs (lncRNAs), which modulate chromatin organization, transcription, and epigenetic states. Although high-resolution structures of lncRNAs are still unavailable due to their size and conformational heterogeneity, biophysical approaches such as atomic force microscopy (AFM) and SAXS, applied for instance to lncRNA MEG3, Braveheart, and XIST, have revealed highly dynamic and modular architectures, suggesting that conformational plasticity is central to their regulatory function (Aguilar et al., 2022; Kim et al., 2020; Uroda et al., 2019).

Finally, a growing number of structured noncoding RNAs of still emerging function are being characterized at high resolution. For example, recent cryo-EM analysis of the bacterial raiA noncoding RNA, which is present in 2,700 bacteria and is as abundant as rRNA and tRNA, has revealed a conserved architecture and suggested mechanisms for its interaction with the translational machinery (Badepally et al., 2024; Haack et al., 2025; He et al., 2026; Kretsch, Wu, et al., 2025). Its structure revealed a novel 3D fold (Badepally et al., 2024), in which stem P1 (formed by the pairing of the 5'- and 3'-ends of the RNA) is crucial for folding. This stem

| Box 3 – Viral RNAs |
|---|
| **Case study 4. tRNA-like structures (TLS)**<br>Key observations:<br>• BMV TLS comprises 3 coaxially stacked helical domains: D1 (helices B1, B2, and C), D2 (helix A, acceptor stem for tyrosylation and the replication initiation site, and helix D), and D3 (helix B3, analogous to the $tRNA^{Tyr}$ anticodon, and helix E).<br>• domain 3 (D3) rotates by ~90º between its free state and the state bound to cognate partner, tyrosyl-tRNA synthetase (TyrRS) (**Figure 4**)<br>• dynamics of domain 3 (D3) consist of formation and deformation of helices B3 and E (**Figure 4**)<br>Organism: *Brome mosaic virus*<br>Number of structures: 3<br>PDB id: 7SAM, 7SC6, and 7SCQ<br>Resolution: 4.3 Å<br>References: (Bonilla et al., 2021) |
| **Case study 5. 5'-proximal regions of Coronaviridae genomic RNAs**<br>Key observations:<br>• flexible arrangement of SL1-5 motifs<br>• SL5 forms a stable T-shaped 4-way junction ($T_m$ 60 °C in MERS-CoV, 63 °C in HKU9-CoV), with two flexible conserved hexaloops (5'-UUYYGU-3') important for protein-RNA and RNA-RNA interaction in the terminal loops of SL5a and SL5b<br>• SL5 element from MERS and Bt-HKU5 viruses showed 3 and 4 distinct conformations, respectively, breaking the coaxial stacking of SL5a and SL5b through swing-like motion around a hinge in SL5a (angular motion of 81º to 84º and 84º to 88º resulting in increased intra-hexaloop distance from 74 to 85 Å and 71 to 85 Å for MERS and BtCoV-HKU5 SL5, respectively) (**Figure 4**)<br>• mutations that disrupt SL5 four-way junction enhance translation by 20-30%, but mutations in SL5c terminal loop affecting SL5c dynamics reduce translation by 45%<br>Organism: *SARS-1, SARS-2, MERS, BtCoV-HKU5, HKU9-CoV, OC43-CoV, RoBat-CoV GCCDC1, HCoV-NL63 and HCoV-229E Coronaviruses*<br>Number of structures: 12<br>PDB id: 8QO2, 8QO5, 8QO4, 8QO3, 8UYS, 8UYP, 8UYK, 8UYL, 8UYM, 8UYE, 8UYG, 8UYJ<br>Resolution: 4.7-9.1 Å<br>References: (de Moura et al., 2024; Jones & Ferre-D'Amare, 2025a; Kretsch et al., 2024) |

has been subsequently harnessed for scaffold-based structure determination (Haack et al., 2025). Recently, structures of distinct raiA subtypes defined a conserved structural core (P3a-P3b, P4, and two pseudoknots) and identified candidate protein-binding partners (KhpA-KhpB) that likely bind to satellite-like peripheral stem loops of the RNA (He et al., 2026).

## *Viral RNAs*

Viral RNA genomes harbor many regulatory and structured RNA elements that are essential for viral infection (Nalewaj & Szabat, 2022). Structural studies have started to reveal the heterogeneous organization of packaged viral RNA genome (Cai et al., 2024) and the dynamics of viral proteins during packaging of the RNA genome in the capsid (Xia et al., 2024). Cryo-EM has also been used to determine structures of viral RNA regulatory elements in complex with host proteins, such as IRES elements bound to ribosomes (Bhattacharjee et al., 2026) or the HIV-1 Rev Response Element (RRE) bound to Rev protein and Crm1 complex, (Fang et al., 2013; Smith et al., 2024).

Meanwhile, the understanding of structure and dynamics of viral RNAs alone is limited. Currently, there are five examples of cryo-EM studies of protein-free viral RNA elements, including the 30 kDa HIV-1 Dimer Initiation Site (DIS) RNA (Zhang et al., 2018), the SARS-CoV-2 frame-shifting stimulatory element (FSE) (Zhang et al., 2020) – which interestingly represents the lowest molecular weight RNA cryo-EM structure in PDB (88 nucleotides, ~28 kDa) –, the FSE element from Rous sarcoma virus (RSV) (Jones & Ferre-D'Amare, 2025b), the tRNA-like motifs (TLS) (**Box 3 – Case study 4**, and **Figure 4**) and 5'-proximal regions of Coronaviridae genomic RNAs (SL1-5) (**Box 3 – Case study 5**, and **Figure 4**).

### *Synthetic RNAs*

Besides naturally occurring RNAs like the ribozymes, riboswitches, and viral RNA motifs described above, synthetic RNAs are also being generated and characterized thanks to computational design or *in vitro* evolution experiments, for applications in nanotechnology. For instance, the ATP-TTR-3 aptamer was designed computationally to specifically bind AMP (Yesselman et al., 2019). Its structure was the first cryo-EM structure determined for a synthetic RNA (Kappel et al., 2020). The structure was solved at 10 Å resolution for AMP-bound and unbound forms of the aptamer and displayed a clothespin-like scaffold. Instead, *in vitro* evolution experiments lead to the emergence of an RNA polymerase ribozyme, which uses trinucleotide triphosphate (pppNNN) and linear or circular RNA templates for RNA replication (Attwater et al., 2018). Its cryo-EM structure revealed functionally important dynamics of this synthetic ribozyme, as highlighted below.

### *Multimeric RNAs*

Both natural and synthetic RNAs have recently been shown – largely thanks to cryo-EM – to be able to form multimeric complexes.

Among biologically occurring multimeric RNAs, the cryo-EM structure of Ornate Large Extremophilic (OLE) RNA was first determined at 2.6-3.1 Å resolution (Kretsch, Wu, et al.,

| BOX 4 – SYNTHETIC RNAS |
|---|
| **Case study 6. RNA polymerase ribozyme** |
| Key observations: |
| • upturned left hand structure with two subunits: 5TU (thumbs, catalytic domain formed by helices P3-P7 and peripheral domain formed by helices P1 and P8-P10); t1 (fingers, helices P1-P3). Kissing loops between P1 (5TU) and J1/2 (t1), and between P7 (5TU) and P3 (t1) confer rigidity |
| • P1 (t1) and P10 (5TU) move towards the active site (P3-P7 in 5TU) by about 30-35° and 20°, respectively (**Figure 5**) |
| • P1-P10 dynamics is regulated by hinge residues located close to an A-anchor motif in P1 and terminal loop and junctions in P10. |
| • P1-P10 dynamics essential for functionally important interactions with the substrate |
| • RNA polymerase ribozyme variants lacking P10 display significantly reduced fidelity |
| Organism: *synthetic RNA* |
| Number of structures: 1 |
| PDB id: 8T2P |
| Resolution: 5.0 Å |
| References: (McRae et al., 2024) |

| Box 5 – Multimeric RNAs |
|---|
| **Case study 7. RNA origami**<br>Key observations:<br>• the '6-helix bundle with a clasp' (6HBC) origami forms a hexagonal arrangement of 6 helices (H1-H6) and adopts distinct conformations during transcription and folding: H6 rotates by 177° around its helical axis and of about 38° between its state immediately after transcription and its state ~10 h post-transcription. This rotation involves remodeling of a kissing loop formed between the two halves of H6 (**Figure 6**)<br>• extensions of 6HBC, such as the '16-helix satellite' (16HS) origami, comprising other synthetic units called '5-helix tile' (5HT) and '6-helix bundle' (6HB) connected by kissing loops, form a Y-shaped antibody-like structure. Individual-particle cryo-electron tomography (IPET) revealed 16 different conformations of three structural modules of 16HS, revealing 5HT-6HB kissing loop rotations of about 150° (**Figure 6**)<br>• in the apo '14-14 traptamer' device, the iSpinach motif is bent by about 68° with respect to the ligand bound state, positioning residues U22 and A45 in the ligand binding pocket and thereby inhibiting the binding of the fluorogenic dye (**Figure 6**).<br>Organism: *synthetic RNAs*<br>Number of structures: 6<br>PDB id: 7PTQ, 7QDU, 7PTS, 7PTK, 7PTL, 8TVZ<br>Resolution: 4.0-5.9 Å<br>References: (McRae et al., 2023; Vallina et al., 2024) |

2025; Puerta-Fernandez et al., 2006; Wang, Xie, Gong, et al., 2025). OLE is a bacterial RNA involved in metal ion homeostasis and cellular adaptations, but its precise molecular mechanism is unclear (Breaker et al., 2023; Wallace et al., 2012). The cryo-EM studies showed OLE's homodimeric architecture and how this RNA exposes to the surface binding sites for its protein partners OapA and OapC.

Second, the cryo-EM structure of another bacterial RNA, which may play a role in plasmid replication and is thus called Area Required for Replication in a Plasmid Of *Fusobacterium* (ARRPOF), was determined at 4.0 Å resolution (Wang, Xie, Gong, et al., 2025). ARRPOF forms a homodimer complex, too, and its dimerization is mediated by kissing loops and intermolecular palindromic interactions.

Third, the cryo-EM structure of Giant, Ornate, Lake- and Lactobacillales-Derived (GOLLD) RNA was solved at 2.5-6.1 Å resolution (Kretsch, Wu, et al., 2025; Wang, Xie, Zhang, et al., 2025; Zhang et al., 2025). GOLLD function is unknown, but its structure revealed a fascinating multimeric nanocage, formed by 14 RNA chains arranged with $D_7$ quaternary symmetry and held together by kissing loops. Remarkably, the 380-Å-diameter cavity within the GOLLD nanocage is sufficiently big to encompass an entire bacterial ribosome.

Fourth, the cryo-EM structure of Rumen-Originating, Ornate, Large (ROOL) RNA, an RNA present in bacterial prophages and phages, was determined at 1.9-3.1 Å resolution (Cousin et al., 2017; Kretsch, Wu, et al., 2025; Wang, Xie, Gong, et al., 2025; Zhang et al., 2025). These structures showed that ROOL forms an octameric nanocage, wherein any one RNA molecule forms 8 bridges with neighboring molecules. These bridges are tertiary A-minor or kissing loop interactions that hold together the ROOL nanocage to form a hollow core large enough to encompass a ribosome, like GOLLD. Structural studies of such multimeric assemblies established how intermolecular tertiary motifs, such as pseudoknots, palindromic

| Box 6 – Chimeric RNAs |
| --- |
| **Case study 8. D43 aptazyme**<br>Key observations:<br>• D43 is a synthetic RNA device composed of a tetracycline-binding aptamer and a hammerhead ribozyme connected by a communication module to control the expression of programmed cell death protein (PD-1) in mammalian cells. The communication module is sandwiched between coaxially stacked RNA helices from the aptamer (helix P1) and the ribozyme (stem II). The communication module is also connected to a non-canonical pseudoknot formed by loop L3 and junctions J1/2 and J2/3, which is part of the tetracycline-binding site.<br>• D43 dynamics are regulated at the nucleotide level, rather than involving large domain motions (RMSD apo vs tetracycline-bound structures = 1.4 Å)<br>• G49•C99 and G48•U100 in the communication module dynamically change conformation: in the apo form they open breaking the coaxial stacking between P1 and stem II and inducing flexibility of the whole D43 (**Figure 7**).<br>Organism: *synthetic RNAs*<br>Number of structures: 4<br>PDB id: 8T5O, 8TKJ, 8TKK, and 8SYK<br>Resolution: 3.0-3.4 Å<br>References: (Stagno et al., 2025) |

base-pairing, stacking and metal-ion coordination, stabilize the RNA in the multimeric assemblies (Huang et al., 2026). These multimeric RNA structures display the potential of RNA to assemble into large scaffold structures with important implications for drug delivery. Because of the potential importance of RNA cages in medicine and nanotechnology, synthetic multimeric RNAs, called RNA origami, have also been developed and characterized by cryo-EM, revealing unique structural dynamics, as highlighted below.

### *Chimeric RNAs*

Finally, both natural and synthetic RNAs have been engineered into fusion constructs, called chimeric RNAs, to combine biological functions or facilitate structural characterization by increasing the size of the target. Using this strategy, cryo-EM structures have been determined, for example, for the Zika virus xrRNA fused to a group I intron (at 4.5 - 5.0 Å resolution) (Langeberg & Kieft, 2023). Strategic use of group II intron as scaffold yielded the best resolution cryo-EM structures of ligand bound/unbound TPP riboswitch or the raiA bacterial ncRNA fused to a group II intron (at 2.5 and 3.0 Å resolution, respectively) (Haack et al., 2025). Negative stain images (but not molecular structure models) were also obtained for a mango-III aptamer, a pre-tRNA, or a primary microRNA (pri-miRNA) fused to ROOL (Ling et al., 2025). Finally, two new *ad hoc* engineered scaffolds with two- and four-fold symmetry, respectively, were used to determine the cryo-EM structures of $tRNA^{Asp}$, Mango-III, quinine and 8-oxoguanine aptamer at 3.7, 3.0, 2.9, and 2.9 Å resolution, respectively (Jones & Ferre-D'Amare, 2026). Interestingly, a completely synthetic chimeric RNA, called the D43 aptazyme, has also been developed to act as a gene regulation device. Its cryo-EM structure revealed peculiar dynamics between its functional modules, as highlighted below.

Although the systems discussed above span RNA classes with very different sequences, structures, and functions, ranging from catalytic, to regulatory, viral, synthetic, and multimeric RNAs, several common principles emerge (**Table 1**). First, RNA dynamics are frequently mediated by the motion of discrete structural modules, such as helices, junctions, pseudoknots, or entire domains, that move through bending, twisting, docking, and undocking motions rather than through complete refolding of the molecule. Second, these conformational changes are directly linked to biological function, enabling substrate recognition, catalysis, ligand sensing, assembly, or long-range communication between distant RNA elements. Finally, the examples discussed here illustrate how cryo-EM uniquely contributes to revealing the structural basis of these motions by visualizing multiple coexisting conformations within the same experimental dataset, providing direct three-dimensional descriptions of dynamic states that are often only indirectly inferred from biochemical and biophysical measurements.

## Challenges inherent to cryo-EM studies of dynamic RNAs and how to overcome them

The case studies highlighted above were successful thanks to specific computational, experimental, and technological developments that helped overcome the intrinsic challenges posed by dynamic RNAs, and particularly challenges in sample design, sample preparation and data acquisition and processing, as discussed below.

### *Challenge 1: Construct design and engineering of dynamic RNAs*

Designing and engineering suitable constructs is an important step before performing a cryo-EM experiment with dynamic RNA molecules and can be based on computational predictions, biochemical information or previously solved structures.

To capture catalytic dynamics of the *Tetrahymena* group I intron, this ribozyme has been engineered to introduce an extended internal guide sequence (IGS), which forms dynamic helices that are important for both splicing steps (**Figure 1**). Moreover, a mutation (G264A) was introduced in the active site to slow down the first step of splicing, and 5'- and 3'-exon like substrates were added to capture different conformations during the second step of splicing (**Figure 1**). Modelling with SimRNA (Boniecki et al., 2016) was also used to guide the structure determination of a previously unseen group I intron conformation involving a pseudoknot and a four-way junction, whose relevance was confirmed biochemically, with mutations.

Informed by the architecture of full-length group II intron from previous structures, folding intermediates of this ribozyme could be obtained by engineering several constructs consisting of different domains of the RNA added in a sequential manner (D1, D1-2, D1-3, D1-4, D1-5, **Figure 2**). Engineering of mutations in an important hinge motif further served to validate the mechanistic folding model (**Figure 2**).

No prior structural information was available for viral tRNA-like structures (TLS), making it difficult to model the atomic coordinates of the wild-type RNA in cryo-EM maps (Bonilla et al., 2021). This prompted the authors to design two constructs of BMV TLS RNAs with extended/truncated helices, with the rationale that local differences between WT and mutant constructs can help in assigning helices. One construct with stem B3 extended and stem C shortened, helped identify these helices in the WT, owing to the local differences. The same strategy was applied with another construct with stem D extended and stem B2 shortened (**Figure 4**). A similar strategy was employed for determining the cryo-EM structures of Coronaviridae SL5 motifs (Kretsch et al., 2024). Here, an engineered extended SL5 construct possessing a helix downstream of SL5 (helix SL6) was designed and characterized by cryo-EM to help identify the 3'-terminal region of SL5 and thus correctly assign the nucleotides in the density (**Figure 4**). Additionally, a distal loop of SL5a was mutated from UUUCGU to GAAA to enable SL5 crystallization *via* formation of novel intermolecular crystal contact, and crystal structure determination in turn facilitated the interpretation of the structural dynamics in the corresponding medium resolution cryo-EM map (Jones & Ferre-D'Amare, 2025a).

*In vitro* evolution experiments were used to select a functionally active and stable RNA polymerase ribozyme (**Figure 5**) (McRae et al., 2024). Synthetic design was also used to generate RNA origami structures, by engineering kissing loops and crossovers using an automated design (ROAD) pipeline (**Figure 6**) (Geary et al., 2021; McRae et al., 2023). Iterative engineering and cryo-EM modeling informed the design of constructs with increasingly improved geometry and shape, and to identify and avoid the formation of kinetic trap during co-transcriptional folding (McRae et al., 2023).

Finally, engineering a stabilizing mutation (U100C) in the communication module allowed better visualization of overall structure and dynamics of the D43 aptazyme in both the holo and apo states (Stagno et al., 2025).

### *Challenge 2: Sample preparation of dynamic RNAs*

After identifying and bioinformatically characterizing the target, the next critical challenging step for the cryo-EM analysis of dynamic RNAs is sample preparation. RNA sample preparation consists of three steps: production, purification and biochemical and biophysical characterization, structural characterization in solution, and cryo-EM grid preparation.

- *Production, purification, and biochemical and biophysical characterization*

RNA synthesis is critical to yield large quantities of homogeneous RNA for grid vitrification and retain the physiological dynamic nature of the RNA target. Recent optimizations of *in vitro* transcription protocols have been crucial to enable successful studies on dynamic RNAs. First, screening of ionic strength (particularly physiological ions such as magnesium and potassium) and pH is crucial because buffer composition affects RNA dynamics and its stability in the vitreous ice (Chillon et al., 2015; Uroda et al., 2020). Recent studies on dynamics RNAs have reported the use of pH in the range from 6.0 to 8.0 and of 1-100 mM magnesium and 10-150 potassium ion concentrations (Chen et al., 2025). It is important to consider that magnesium ions are essential for RNA folding, ligand recognition, and ribozyme activity, but high levels of magnesium may induce RNA particle oligomerization and aggregation, exacerbating a common challenge of dynamic RNAs (Chen et al., 2025). Second, purification strategies also affect RNA dynamics and its folding into physiological, functional structures. Two strategies have been successfully used for studying dynamic RNAs, namely denaturing and non-denaturing purification protocols (Chillon et al., 2015; Nilsen, 2013). Purification using denaturing acrylamide gels and refolding has been used for cryo-EM studies of *Tetrahymena* group I intron dynamics (Li et al., 2023; Zhang et al., 2023) (**Figure 1**), TLS and SL5 viral RNA elements (Bonilla et al., 2021; Kretsch et al., 2024), and the cobalamin riboswitch (Ding, Deme, et al., 2023). In this procedure, the temperature at which refolding and reconstitution with physiological ions occur is crucial. For instance, *Tetrahymena* group I intron folding intermediates have been obtained at low temperature (25 °C), whereas the native states have been obtained at higher temperature (50 °C) (Bonilla et al., 2022; S. Li et al., 2022). Non-denaturing purification has been utilized for cryo-EM studies of group II intron folding intermediates (Jadhav et al., 2025), RNA polymerase ribozyme (McRae et al., 2024), group I intron (Luo et al., 2023), RNA origami scaffolds (McRae et al., 2023; Sampedro Vallina et al., 2023; Vallina et al., 2024), RNA D43 aptazyme (Stagno et al., 2025) and raiA ncRNA (Badepally et al., 2024).

Followed by purification, biochemical and biophysical characterization is necessary to confirm that the target RNA is pure and functionally active, to obtain initial insights into its dynamics, and to exclude the formation of non-specific aggregates or unfolded molecules (Chen et al., 2025). Studies on engineered ribozymes have confirmed activity using enzymatic assays (**Figures 1-2**). Furthermore, to confirm homogeneity, studies on group II intron folding intermediates, RNA origami (5HT, 6HB, 6HBC, 16HS, and traptamer), and the D43 aptazyme have used gel electrophoresis and size exclusion chromatography (SEC) (Jadhav & Marcia, 2025; McRae et al., 2023; Stagno et al., 2025). It is worth noting that in some cases analytical gel electrophoresis shows high sample homogeneity (i.e. BMV TLS, **Figure 4**) while in other cases it displays multiple bands (i.e. group I and II introns, **Figures 1-2**, and D43 aptazyme, **Figure 7**), but these samples are nonetheless suitable for cryo-EM investigation. Studies on group II intron folding intermediates have additionally used mass photometry (MP) and size exclusion chromatography coupled to multiangle laser light scattering (SEC-MALLS) to identify unfolded or aggregated forms of the target and to gain information regarding molecular weight and hydrodynamic properties of the target (Jadhav & Marcia, 2025). These latter techniques require significantly different amounts of RNA for analysis, i.e. micromolar concentrations for SEC-MALLS and nanomolar concentrations for MP. Thus, comparison of how the RNA behaves in these different techniques is particularly informative to judge if concentration plays a role in determining RNA homogeneity.

Finally, biochemical probing can also support studies on RNA structural dynamics, specifically maps the secondary structure of the RNA target and thus facilitates model building in the cryo-EM maps. For instance, the Ribosolve pipeline, which integrates secondary structure probing using the so-called "mutate-and-map read out through next-generation sequencing" approach (M2-seq), is now largely used in RNA cryo-EM studies (Cheng et al., 2017; Kappel et al., 2020). Such approach was used to elucidate the structure of the SL5 element of *Coronaviridae* (**Figure 4**) (Kretsch et al., 2024). Chemical probing studies has also revealed the deformation of specific tertiary motifs during the folding process of group II introns (Waldsich & Pyle, 2008). Now, the structural basis behind such deformation and its implications for group II intron dynamics and folding has been confirmed and rationalized by cryo-EM studies on group II intron folding intermediates (**Figure 2**) (Jadhav et al., 2025).

- In-Solution *RNA characterization*

After purification and before cryo-EM imaging, visualizing RNA dynamics in solution is beneficial to gain initial insights into the range of conformational changes that the target undergoes. Three techniques have been used in recent cryo-EM studies of dynamic RNAs, namely SAXS, NMR and AFM.

First, SAXS has been used to study group II intron folding intermediates (Jadhav & Marcia, 2025), 6HB RNA origami nanostructures (McRae et al., 2023), and the D43 aptazyme (**Figure 7**) (Stagno et al., 2025). SAXS provides insights into RNA flexibility, hydrodynamic parameters like the radius of gyration ($R_g$) and the maximum pairwise interatomic distance ($D_{max}$), and the overall low-resolution shape of the molecule, which can later support RNA structure modeling into the cryo-EM maps (Chen & Pollack, 2016; Kim et al., 2020). In addition, SAXS informs about the effect of ions or ligands on RNA conformation (Pollack, 2011). For instance, the folding of domain 3 of group II introns, which forms a characteristic 'S-turn' motif (Marcia & Pyle, 2014), is significantly impacted by the ionic composition of the buffer and SAXS studies could capture this dependence. In the absence of potassium ions, group II intron domain 3 is elongated, but in the presence of potassium ions it docks onto the globular scaffold of the intron (Jadhav & Marcia, 2025). Finally, SAXS helps characterize conformational ensemble states of the target and suggests if the cryo-EM data have captured all or only some of the conformational states explored by the target in solution. For instance, the back-calculated

SAXS curves of apo- and holo-D43 aptazyme cryo-EM structures differ from the experimental scattering curves (**Figure 7**). The deviation is greater for the apo-state compared to the holo-D43 aptazyme. Additionally, cryo-EM structure-derived curves are similar to each other compared to the comparison of SAXS experimental curves. This discrepancy suggests that there are structural differences between the apo- and holo-D43 aptazyme (**Figure 7**) (Stagno et al., 2025).

Second, NMR provides dynamics insights at the single nucleotide level, i.e. changes in base pairing or base stacking patterns, and it was used to support cryo-EM studies on the dynamic D43 aptazyme (**Figure 7**) (Stagno et al., 2025). Such conformational changes may be difficult to visualize in the cryo-EM maps, for instance due to modest resolution. In the case of the D43 aptazyme, NMR helped define the conformation of nucleobases from the communication module to transduce the signal from the aptamer to the ribozyme module (**Figure 7**) (Stagno et al., 2025).

AFM together with SAXS and isothermal titration calorimetry (ITC) was used in support of cryo-EM studies on the dynamics of the cobalamin riboswitch (Ding, Deme, et al., 2023; Ding, Lee, et al., 2023). AFM captures the 3D shape and contour information of the target RNA immobilized on mica surfaces. For the cobalamin riboswitch, AFM specifically helped identifying distinct orientation of RNA helices and the dimerization of the RNA (**Figure 3**) (Ding, Lee, et al., 2023). Finally, studies, a combination of dry and in solution AFM showed the dynamics and organization of the SL1-5 motifs of the 5'-proximal region of *Coronaviridae* genomes. Additionally, AFM integration with thermal shift assays further confirmed the dynamic nature and independent degree of flexibility of SL1-5 (de Moura et al., 2024). Temperature-dependent SHAPE experiments further provided nucleotide level information on temperature induced secondary and tertiary interactions (de Moura et al., 2024).

- *Cryo-EM grid preparation*

After production, purification and biochemical and biophysical characterization, the RNA target is ready for cryo-EM investigation of its dynamics, at high resolution. The first step to prepare the sample for cryo-EM is grid preparation. Recent studies on dynamic RNAs have shown that two steps in grid preparation are important, namely inspection by negative staining followed by grid vitrification.

First, negative staining is rarely used nowadays, considering the easier accessibility and higher throughput of vitrification approaches. But studies on dynamic RNAs have shown that negative staining can still provide useful quick insights into the homogeneity of the target, its stability, and its behavior on the EM grids. For instance, negative staining procedure have been used to visualize dynamics of the 16HS RNA origami and of group II intron folding intermediates (**Figure 6**) (Jadhav & Marcia, 2025; McRae et al., 2023). Initial views into negative staining images of RNAs may display compact, flexible, or unfolded thread-like molecules of the RNA sample. Depending on the RNA behavior, one may revisit RNA purification before proceeding with vitrification trials.

Second, grid vitrification is affected by several factors. The choice of the grid type is critical because it is one of the factors that determines the thickness of the vitreous ice. Different sizes of the holes in holey carbon and the metal support, typically copper or gold, can yield different ice thickness after plunge freezing and affect beam-induced particle motion, which consequently affects the resolution of the data (Russo & Passmore, 2014). The cobalamin riboswitch (Ding, Deme, et al., 2023), the D43 aptazyme (Stagno et al., 2025), the traptamer RNA origami (Vallina et al., 2024), the 16HS and 6HBC RNA origami (McRae et al., 2023), and the RNA polymerase ribozyme (McRae et al., 2024) have been vitrified using 300 mesh 1.2/1.3 Au grids. Group II folding intermediates have been vitrified using 300 mesh 1.2/1.3 Cu grids (Jadhav et al., 2025; Jadhav & Marcia, 2025). Group I intron samples have been vitrified using 200 mesh 2/1 Au or Cu grids (Li et al., 2023; Luo et al., 2023; Zhang et al., 2023). A 72 kDa raiA construct has been vitrified using UltraAuFoil (He et al., 2026) and Quantifoil R2/2, 200 mesh Cu grids (Badepally et al., 2024). To improve the raiA particle distribution, experimentalist used 0.05% Igepal CA-630 (He et al., 2026). These examples show that although the choice of the size of the holey carbon holes should in principle be directed by the length of the target RNA, 1.2 µm holey carbon layers have been used successfully for the grid vitrification of RNA samples ranging from 100 kDa ribozymes to megadalton-large RNA nanocage structures (Haack et al., 2025; Kretsch, Wu, et al., 2025; Ling et al., 2025).

### *Challenge 3: Cryo-EM data collection and processing of dynamic RNAs*

After successfully depositing the target RNA on cryo-EM grids, the next step towards structure determination consists of data collection and processing. Here, challenges specific to dynamic RNAs can be identified at the screening, data collection, data processing, and structure reconstruction steps.

- *Cryo-EM grid screening*

Grid screening is necessary to identify grids with optimal particle distribution and density throughout grid squares. Dynamic RNA particles often do not behave homogenously. Reports on which grid squares are most successful for structure determination of dynamic RNAs are rare, but recent studies on group II intron folding intermediates showed that dynamic RNAs prefer thicker vitreous ice squares compared to thinner ones (Jadhav & Marcia, 2025), and this may be a generalizable feature typical of dynamic RNAs. Thus, dynamic RNAs display optimal distribution and well-folded particles in the smaller squares, which harbor thicker ice (**Figure 2**) (Jadhav & Marcia, 2025).

- *Cryo-EM data collection*

The thicker ice required by dynamic RNAs poses a challenge in data collection for two reasons.

First, thick ice squares can hamper localization of holey carbon holes. This problem can now be efficiently circumvented thanks to recent technological advances, i.e. direct alignment approaches, wherein an individual hole is centered just before image acquisition, or the newer "junk detector" tool in cryoSPARC, which can exclude micrograph images collected on carbon. However, aberration-free image shift (AFIS), which is beneficial to enable faster and larger data acquisitions, is not efficient when collecting on grids with thicker ice. Should AFIS be utilized for data collection on dynamic RNA samples, then the user should pay particular attention to exclude from the dataset images collected on the carbon support instead of in the holes, before downstream data processing.

Second, thick ice would typically require higher acquisition dose and would thus risk to induce particle radiation damage or more pronounced beam-induced motions. But the high 'Z' number of phosphorus, which is abundant in RNA, increases RNA's overall electron scattering potential and allows using proportionally lower electron dose for data collection on RNA vs proteins. While a systematic study on the effect of electron dose on image quality has been performed only for RNA origami nanostructures (McRae et al., 2023), an empirical compromise used in most other recent studies on dynamic RNAs is that the same dose that is suitable for proteins in thin ice (i.e. 40-60 $e^-/Å^2$) is also suitable for dynamic RNA in thicker ice. For instance, an electron dose of 60 $e^-/Å^2$ has been used to image the TLS element of BMV (Bonilla et al., 2021), the RNA polymerase ribozyme (McRae et al., 2024), and group I introns (Luo et al., 2023). An electron dose of 50 $e^-/Å^2$ has instead been applied to image other group I intron samples (Li et al., 2023; Zhang et al., 2023), the cobalamin riboswitch (Ding, Deme, et al., 2023), and the D43 aptazyme (Stagno et al., 2025). A lower electron dose of 40 $e^-/Å^2$ has been used to acquire movies of group II intron folding intermediates (Jadhav et al., 2025).

- *Cryo-EM particle picking*

During data processing, particle picking needs to ensure that aggregated or unfolded particles, which pose a particular challenge for dynamic RNAs (see above), are excluded from the dataset. To facilitate particle picking, the choice of an appropriate box size, i.e. a box significantly larger than the expected particle dimensions is beneficial, because it helps to distinguish between folded particles and unfolded thread-like particles (Kretsch et al., 2024). Of note, RNAs that possess both structured and flexible regions may appear as a comet on raw micrographs, where the head corresponds to the structured region of the RNA and the tail to its flexible domains (**Figure 2**) (Jadhav & Marcia, 2025). For RNAs smaller than 100 kDa, it is beneficial to perform manual particle picking on high-defocus micrographs for template picking (Badepally et al., 2024). Both *CryoSPARC* and EMAN2 have been used successfully for particle picking in recent studies on dynamic RNAs (Punjani et al., 2017; Tang et al., 2007). For instance, group I intron samples were processed using pipeline involving motion correction (MotionCor2 (Zheng et al., 2017)), CTF estimation (CTFFIND4 (Rohou & Grigorieff, 2015)), particle picking with EMAN2 (Tang et al., 2007), 2D classification in Relion (Scheres, 2012) and 3D classification and 3DVA analysis in *CryoSPARC* (Punjani & Fleet, 2021; Punjani et al., 2017). Instead, studies on the BMV TLS element (Bonilla et al., 2021), RNA polymerase ribozyme (McRae et al., 2024), the D43 aptazyme (Stagno et al., 2025), RNA origami nanostructures (McRae et al., 2023), raiA ncRNA (Badepally et al., 2024), and the SL5 viral element (Kretsch et al., 2024) used *CryoSPARC* for the entire data processing pipeline. Cryo-EM structures of group II intron folding intermediates were acquired using *CryoSPARC* and JANAS (Jadhav et al., 2025; Maiorca et al., 2026). Cobalamin riboswitch study used SIMPLE software (Caesar et al., 2020) for micrograph pre-processing and *CryoSPARC* and Relion for 3D reconstruction (Ding, Deme, et al., 2023).

- *Cryo-EM 3D structure reconstruction*

Besides particle picking, another important step in data processing of dynamic RNAs is 3D reconstruction and the visualization of the target's conformational heterogeneity.

The first step in 3D reconstruction is *ab initio* reconstruction. *Ab initio* reconstruction provides the initial map of the RNA and therefore affects the accuracy and quality of the final, refined maps (Ma et al., 2022). Once an initial map is acquired, refinement consists of homogenous refinement steps to improve the resolution of the initial map, and heterogeneous refinement steps to gain insights into the heterogeneity of the target. Biological macromolecular structures display two types of heterogeneity, compositional and conformational heterogeneity. Compositional heterogeneity exists due to distinct molecular composition of the same target. In contrast, conformational heterogeneity, which corresponds to the presence of discrete or continuous conformational states, arises from the inherent motion of different domains within the same macromolecular complex. Resolving both types of heterogeneity is essential to determine the high-resolution structure and gain functional understanding into dynamic RNAs, but this is a challenge, partly due to unique signal to noise ratio associated with each cryo-EM sample and dataset which makes it difficult to develop algorithms that can process all cryo-EM samples (Kimanius & Schwab, 2024). Recent studies on dynamic RNAs have successfully used three different approaches to resolve RNA compositional and conformational heterogeneity. 3D classification in Relion (Kimanius et al., 2016) or *CryoSPARC* (Punjani et al., 2017) has been successfully used to discern distinct conformations of the cobalamin riboswitch (Ding, Deme, et al., 2023). 3D classification in Relion has also been used to reveal the dynamics that govern the progression of group I intron throughout its self-splicing reaction (Luo et al., 2023). 3DVA (Punjani & Fleet, 2021) has been successfully applied to gain functional insights into the dynamics of group I introns (Li et al., 2023), group II intron folding intermediates (Jadhav et al., 2025), the viral SL5 and TLS elements (**Figure 4**) (Bonilla & Kieft, 2022; Bonilla et al., 2021; Kretsch et al., 2024; Yang et al., 2025), 5HT and 6HBC RNA origami structures (McRae et al., 2023) and the RNA polymerase ribozyme (McRae et al., 2024). Finally, a recent algorithm still under development (JANAS), which helps identify discrete low occupancy conformational states, has also been successfully employed for resolving group II intron dynamics (Maiorca et al., 2026).

Other approaches to resolve compositional and conformational heterogeneity have been developed recently, albeit not having been used in studies on RNA dynamics yet, i.e. deep-learning approaches such as CryoDRGN (Zhong et al., 2021) and e2gmm (Chen & Ludtke, 2021; Chen et al., 2023). These methods, along with 3DVA, have substantially expanded the ability of cryo-EM to characterize conformational heterogeneity and thus have the potential to make a substantial impact also on processing heterogeneity in RNA cryo-EM datasets, but important challenges remain in relating experimentally observed ensembles to underlying RNA energy landscapes (Zhong et al., 2021). Single-particle cryo-EM captures a distribution of vitrified molecular snapshots rather than the dynamic transitions connecting them, limiting direct access to kinetic information and transition rates. Moreover, dimensionality-reduction approaches necessarily project complex conformational ensembles onto low-dimensional representations, and the resulting landscapes should therefore be interpreted as approximations of experimentally sampled structural variability rather than direct measurements of free-energy surfaces (Zhong et al., 2021). Experimental factors such as limited sampling of rare states, preferred orientations, classification bias, and perturbations associated with vitrification may further influence the observed state distributions (Zhong et al., 2021). Consequently, cryo-EM-derived conformational ensembles are most powerful when integrated with complementary approaches, including molecular dynamics simulations, smFRET, SAXS, and solution-based structure probing, which together can provide a more complete description of RNA energy landscapes and conformational transitions (see also “Challenge 2: Sample preparation of dynamic RNAs”, above).

## Conclusions and Future Perspectives

The study of RNA has long been constrained by static representations of inherently dynamic molecules. Here, we review recent advances in the field that collectively point to a fundamental transition: RNA structural biology is moving beyond the determination of individual conformations toward the reconstruction of dynamic ensembles and, ultimately, energy landscapes. Cryo-EM has emerged as a central driver of this transition by enabling direct visualization of both discrete structural states and continuous conformational variability across a broad spectrum of RNA systems.

RNA function is encoded not in a single structure, but in the accessibility, interconversion, and regulation of multiple conformational states. Across diverse classes of RNAs – including ribozymes, riboswitches, viral elements, and synthetic assemblies – cryo-EM is now enabling to describe the dynamics that enables RNA function. The case studies discussed in this review illustrate how cryo-EM can capture RNA dynamics states and the motions that connect them, revealing mechanisms such as hinge-mediated domain rearrangements, ligand-dependent switching, and cooperative assembly processes. Remarkably, these seemingly diverse phenomena often arise from a repertoire of recurrent structural motions, including helix rotations, bending around multi-way junctions, remodeling of tertiary contacts, and transitions between docked and undocked states, suggesting that defined sets of structural mechanisms underpin a wide range of RNA functions (**Table 1**). Across all RNA classes, these motions regulate the accessibility of catalytic centers, ligand-binding pockets, protein-interaction surfaces, or assembly interfaces, revealing a common relationship between conformational dynamics and biological function.

Importantly, these insights are increasingly quantitative, providing spatial and mechanistic detail that links structural transitions directly to functional outcomes.

At the same time, these advances highlight the experimental and computational challenges that must be overcome to fully realize this paradigm. The intrinsic flexibility, heterogeneity, and sensitivity of RNA impose stringent requirements on construct design, sample preparation, vitrification, and data analysis. Recent progress in these areas – including improved RNA engineering strategies, optimized vitrification approaches, and powerful computational tools for disentangling conformational heterogeneity – has been essential for extending cryo-EM to dynamic and previously intractable RNA targets. Continued innovation along these axes will be critical for pushing the limits of resolution, sensitivity, and interpretability.

Importantly, cryo-EM occupies a unique position within the toolkit used to study RNA dynamics. While approaches such as smFRET, NMR, SAXS, chemical probing, and molecular simulations provide detailed information on kinetics, local structural changes, or conformational ensembles, cryo-EM directly visualizes the three-dimensional architecture of individual states and increasingly the structural trajectories connecting them. The complementary strengths of these approaches are therefore most fully realized when integrated within a common framework for describing RNA energy landscapes.

Looking forward, the most impactful developments are likely to arise from strengthening the integration of cryo-EM with complementary approaches. Coupling cryo-EM with molecular dynamics simulations, chemical probing, fluorescence microscopy, and other solution-based techniques is starting to make it possible to construct experimentally grounded conformational ensembles that bridge static structures and dynamic behavior. For instance, using fluorescent tags to localize specific RNAs *in situ* (i.e., based on CRISPR/Cas13a or MS2/PP7 systems) enables correlative light and electron microscopy studies linking the dynamics observed *in vitro* to how RNAs behave inside living cells (Cao et al., 2022; Hampton et al., 2017; Moser et al., 2019; Park et al., 2014). Furthermore, the observation of RNA dynamics through cryo-EM synergizes with serial femtosecond crystallography (SFX) which allows the visualization of ultrafast structural intermediates (Chapman et al., 2011; Stagno et al., 2017). In this context, advances in ultrafast vitrification hold the promise of capturing transient intermediates along functional pathways, bringing structural biology closer to a true "molecular cinematography" of RNA processes. For instance, Chameleon provides blot-free self-wicking, ultra-rapid vitrification, and automated sample handling (Levitz et al., 2022), while time-resolved cryo-plunger, Easy-grid, microfluidic chips or micro-sprayers allow for on-the-spot mixing and spray-freezing (Garg et al., 2025; Torino et al., 2023). These approaches reduce RNA denaturation and structural disturbance and increase the chance to capture short-lived states. Computational advances in the analysis of RNA sequences and in the prediction of RNA 3D structures are also starting to play a transformative role. For now, recent CASP competitions (Kretsch et al., 2026) demonstrated that physics-based models, like Vfold (J. Li et al., 2022) and SimRNA (Boniecki et al., 2016) are still generally more accurate in predicting RNA 3D structures *de novo* than AI-based ones, like AlphaFold3 (Abramson et al., 2024), RNAnneal (Herron et al., 2026), DynaRNA (Li et al., 2025), RoseTTAFold2NA (Baek et al., 2024). But these same CASP competitions are also promisingly showing that AI-based RNA structure prediction algorithms can powerfully integrate the prediction of RNA conformational ensembles (Dube et al., 2026) and of dynamic solvent shells surrounding RNA structures, especially when integrated with SAXS, outperforming MD predictions (Kretsch, Posani, et al., 2025; Patt et al., 2025).

But AI is not only improving RNA structural predictions. It is also starting to make an impact on RNA cryo-EM data analysis. AI-driven particle picking and classification are getting better at handling extreme heterogeneity, automatically telling folded cores apart from flexible tails and spotting rare but functionally important states in huge datasets. We can expect that next-gen algorithms will further blend the strengths of tools like cryoDRGN, 3DVA, and JANAS enabling RNA conformational heterogeneity to be represented as continuous structural landscapes rather than collections of discrete states (Maiorca et al., 2026; Punjani & Fleet, 2021; Zhong et al., 2021). These approaches increasingly allow coordinated RNA motions and low-population intermediates to be identified directly from particle images, providing a richer description of conformational ensembles. Tighter integration of cryo-EM heterogeneity analysis with MD simulations, enhanced-sampling techniques, and experimental restraints should help bridge the gap between experimentally observed conformational distributions and underlying energy landscapes. This combination will help reveal exactly how solvent, ions, and structural fluctuations drive function and improve the modeling of flexible or poorly resolved regions (Bernetti & Bussi, 2023; Languin-Cattoen & Bussi, 2026; Wlodarski et al., 2024). More broadly, these advances will overcome the current limitations of cryo-EM, i.e. that cryo-EM captures snapshots of molecules frozen at one instant, not the actual transitions between states and that the relationship between particle abundance and their actual free energy is complicated by experimental biases, so that cryo-EM-derived occupancies do not correspond to quantitative equilibrium populations. By combining cryo-EM-derived conformational distributions with AI-based heterogeneity analysis and physics-based simulations, it may become possible to reconstruct experimentally grounded RNA energy landscapes rather than merely catalog individual conformational states. Such integrative frameworks promise a more quantitative understanding of how RNAs navigate complex conformational ensembles during folding, regulation, and catalysis.

Ultimately, the emerging dynamic view of RNA has important implications beyond fundamental biology. A detailed understanding of RNA conformational landscapes informs the development of RNA-targeted therapeutics by revealing transient binding pockets and regulatory states that can be selectively stabilized or disrupted. Likewise, the ability to design and control RNA dynamics expands the toolkit of RNA nanotechnology, enabling the creation of responsive molecular devices with programmable behavior. In this context, cryo-EM is poised to play a central role, providing the structural framework needed to connect sequence, structure, dynamics, and function.

In summary, the convergence of methodological innovation and conceptual advances is redefining how we think about RNA. By linking structural snapshots into coherent dynamic models, cryo-EM is transforming our understanding of RNA from static architecture to adaptive, functional motion. This shift toward a dynamic, ensemble-based perspective not only resolves long-standing challenges in RNA biology but also opens new avenues for discovery at the interface of structural biology, biophysics, and medicine.

## Acknowledgements

We thank all members of the Marcia lab for helpful discussion. Work in the Marcia lab is partly funded by the Swedish National Research Council (VR, 2024-04107), by the HORIZON-MSCA-2023-DN-01 action (project: TargetRNA, n. 101168667), by Cancerfonden (project 25 4360 Pj), by SciLifeLab (Strategic initiatives and technology development call 2025), and by the Italian Association for Cancer Research (AIRC, IG 28746).

## Author contribution statement

MM conceived and supervised the work and obtained funding; MM, SJ, SS, and QS analyzed the structures; SJ and MM interpreted the data with contribution of all authors; SJ and MM produced the first draft of the manuscript; all authors approved the final version of the manuscript.

## Competing interest statement

The authors declare no competing interests.

## Figures and legends

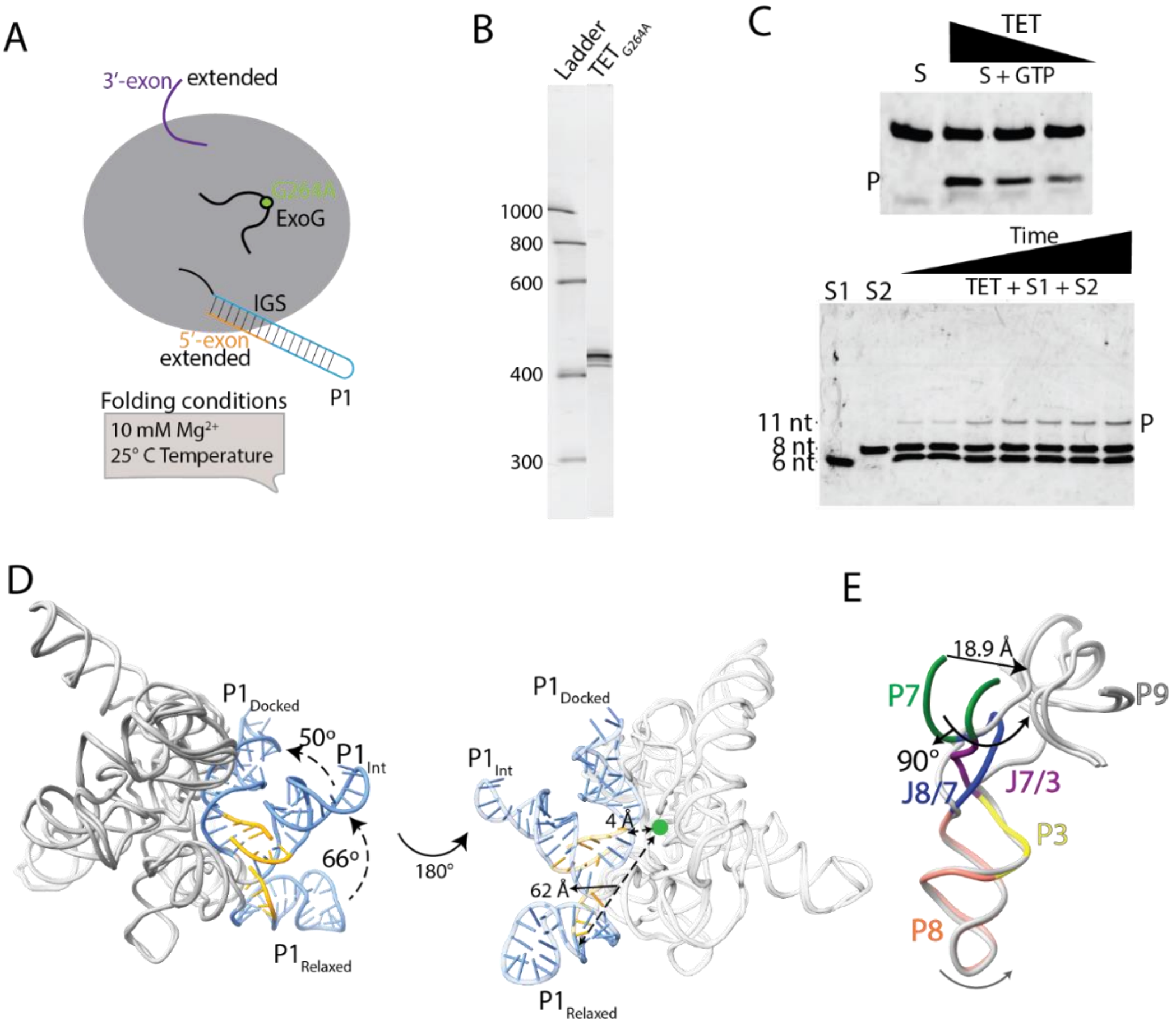

**Figure 1. *Ad hoc* construct design to slow down functional dynamics enabled the capturing of group I intron splicing and folding intermediates. (A)** Schematic representation of group I introns highlighting the engineering strategies that enabled cryo-EM investigation of catalytic intermediates, i.e. insertion of IGS, addition of 5'- and 3'-splice analogues, and active site mutations (G264A). Folding intermediates have been obtained by modulating the refolding conditions, as indicated in the inset (concentration of magnesium, $Mg^{2+}$, and temperature, T). **(B)** Representative analytical PAGE gel of the G264A group I intron sample ($TET_{G264A}$) used for cryoEM. **(C)** PAGE gels confirming activity of the IGS-engineered group I intron. The top gel displays the first step of splicing reaction, the bottom gel the second step of splicing reaction. S, S1 and S2 indicate fluorescently labeled substrates, P indicates the product of the reaction, TET indicates the intron. **(D)** Superposition of co-transcriptionally folded group I intron structures formed in the first step of splicing, and displaying the P1 helix (depicted in orange, for the 5'-exon moiety, and blue, for the intron 5'-end moiety) in the relaxed (PDB id: 8HD6), intermediate (abbreviated as Int, PDB id: 8HD7), and docked (PDB id: 8I7N) states. The distance between the 5'-splice site and the G-binding site (in green) in depicted with dashed black line for the relaxed (62 Å) and docked (4 Å) state. The left and right subpanels are rotated 180° degree with respect to one another. **(E)** Superposition of the misfolded (M, PDB id: 7XSK) intermediate and native (N, PDB id: 7XSN, in grey) states of group I intron. For the M-state, the P7 helix is in green, the P8 helix is in salmon, the P3 helix is in yellow, the J7/8 helix is in dark blue, the J7/3 helix is in purple and the P9 helix is in grey. Panel B has been adapted from Extended Data Figure 7 panel A from (Luo et al., 2023). Top panel C was reprinted with permission from Supplementary Figure 1 panel A from Zhang, X., et al. (2023), Nucleic Acids Research, which was published under a Creative Commons Attribution-Non-Commercial License (https://creativecommons.org/licenses/by-nc/4.0/). Further reproductions must adhere to the terms of this license (Zhang et al., 2023). Bottom panel C is reproduced from Supplementary Figure 1, panel A from Li, S., et al. (Li et al., 2023).

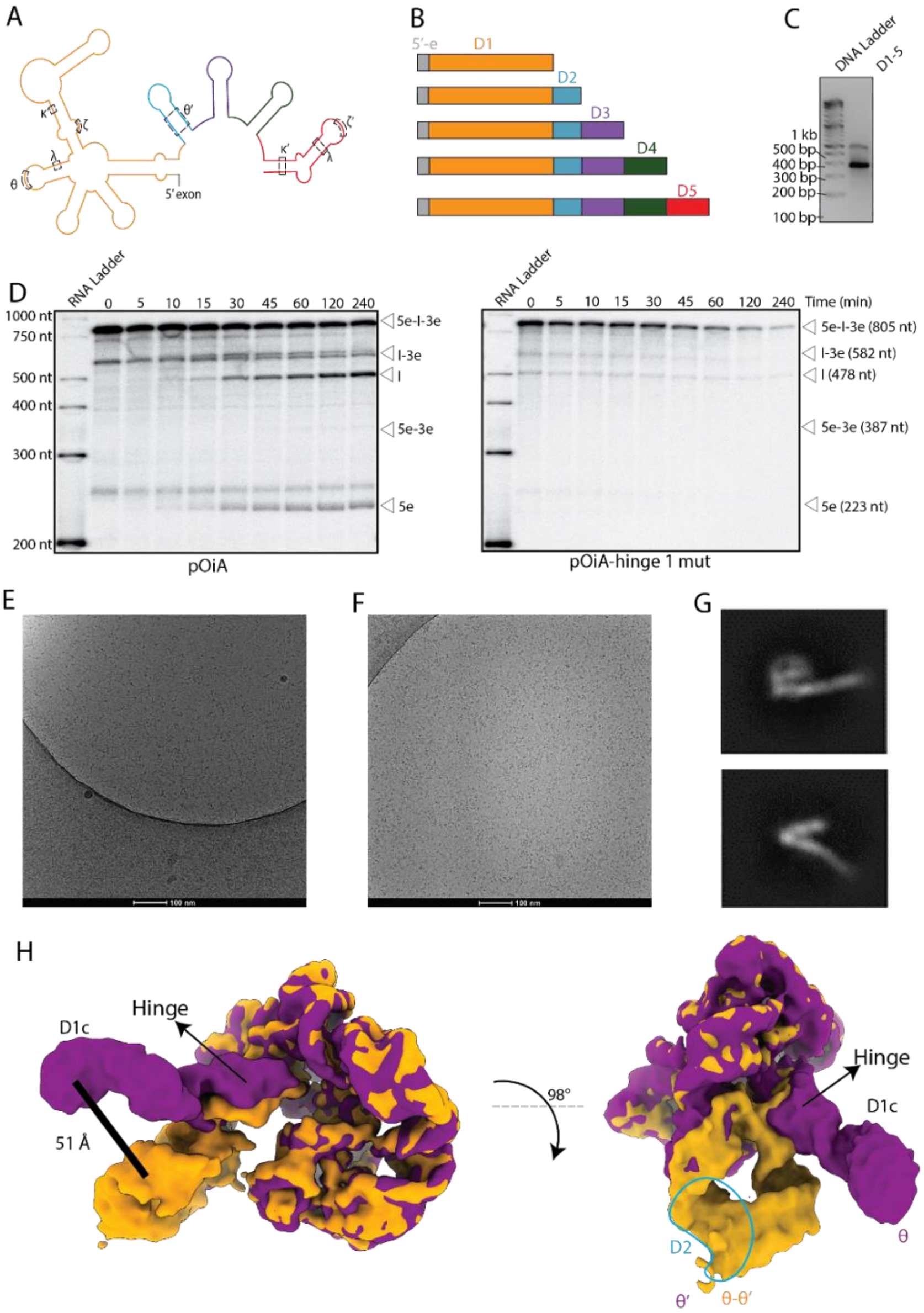

**Figure 2. Construct engineering and optimization of vitrification conditions enabled the visualization of group II intron folding intermediates. (A)** Secondary structure map of the group II intron used for visualizing folding intermediates. The map is colored by domains. Black dashed squares indicate tertiary structure motifs (θ-θ’, κ, ζ, λ) that are flexible in both the cryo-EM maps and in previous chemical probing studies of folding intermediates. **(B)** Constructs used for cryo-EM investigation. The color code of each domain is the same as in panel A. **(C)** PAGE gel showing the purity and homogeneity of a representative construct. **(D)** PAGE gels comparing the activity of wild type and mutant introns, used to validate the folding model. **(E)** Micrographs of the D1-3 construct acquired in suboptimal grid vitrification conditions (31 μM concentration, -3 force, 1 sec blotting) and displaying unfolded, aggregated particles. **(F)** Micrographs of the D1-3 construct acquired in optimal grid vitrification conditions (30 μM concentration, -10 force, 2 sec blotting) and displaying homogenously distributed particles. **(G)** 2D classes showing comet-like particles formed by dynamic group II intron constructs. **(H)** Superposition of different conformations of the D1-3 folding intermediate, highlighting the D1c helix and the hinge motif that control its dynamics. The θ and θ’ nucleotides are also labeled, showing that the θ-θ’ interaction is dynamic during group II intron folding in line with the chemical probing data (see panel A and the main text). The two subpanels are rotated by 98° with respect to one another. Panel C is reproduced from Figure 1 panel B from Jadhav, S. and M. Marcia (Jadhav & Marcia, 2025). Panel D is reproduced from Figure 8 panel B of Jadhav, S., et al. (Jadhav et al., 2025). Panel G is reproduced from Figure 6 panel A from Jadhav, S. and M. Marcia (Jadhav & Marcia, 2025).

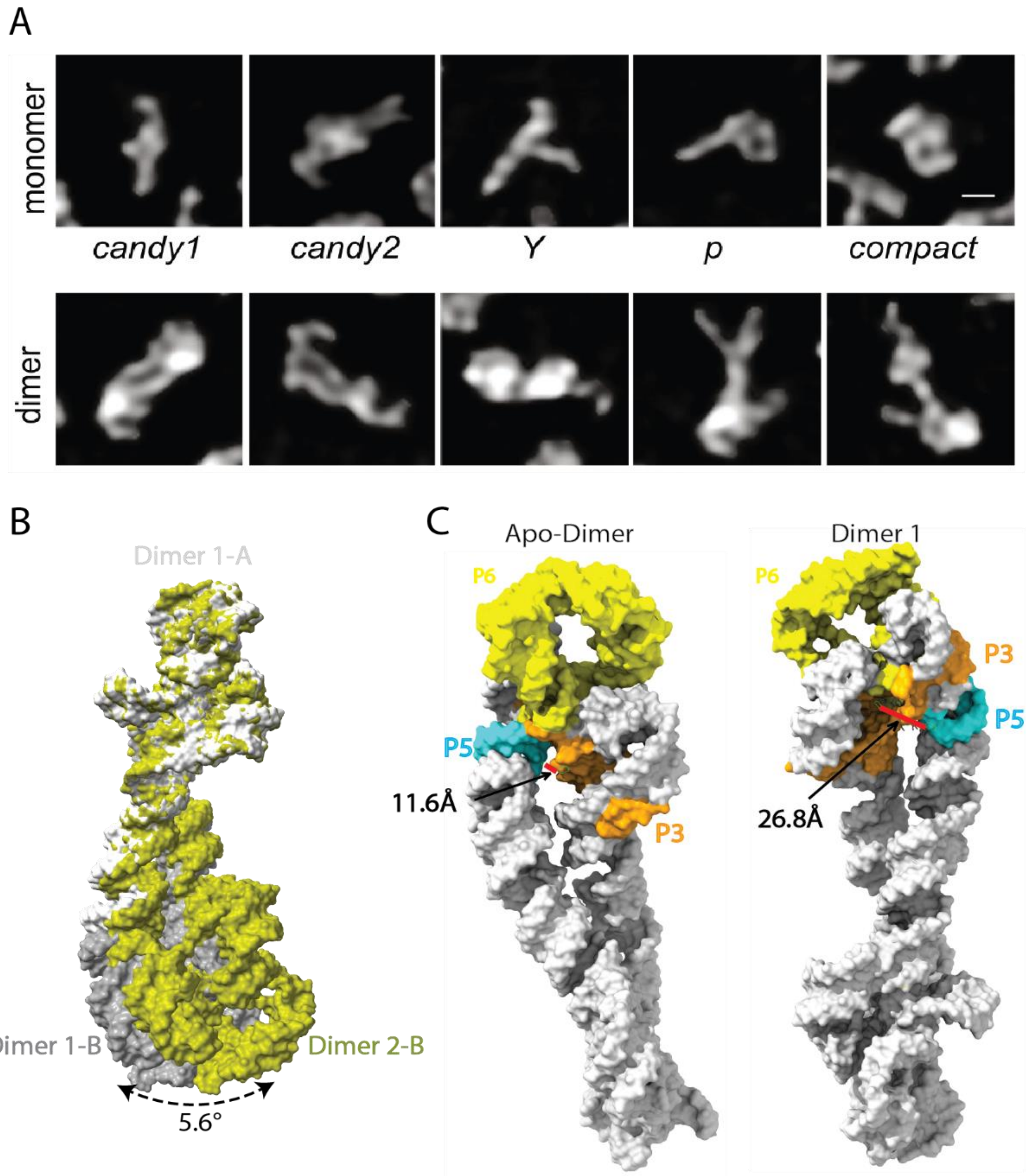

**Figure 3. Integration of cryo-EM and AFM data enabled the visualization of cobalamin riboswitch dynamics between its apo- and its holo-states. (A)** High resolution AFM images of rCbl in monomeric and dimeric states. Dimeric and monomeric states have been classified on the basis of shape. **(B)** Superposition of dimer 1 (PDB ID: 8SA2, shown as surface representation in grey) and dimer 2 (PDB ID:8SA3, in olive). The black dotted arrow indicates the motion of chain B in dimer 2 with respect to dimer 1. **(C)** Comparison of the apo (PDB ID: 8SA6, shown as surface representation on the left) and the holo dimeric forms of the riboswitch (dimer 1, PDB ID: 8SA2, shown as surface representation on the right). The distances between phosphorus atoms of G74 (in the P5 domain; in cyan) and C165 (in the P3 domain; in orange) in both structures are displayed as black lines and labeled (in Å) to quantify the movement of the P6 domain (in yellow) between the two states. Panel A is reproduced from Figure 1 panels C and D of Ding, J., et al. (Ding, Lee, et al., 2023).

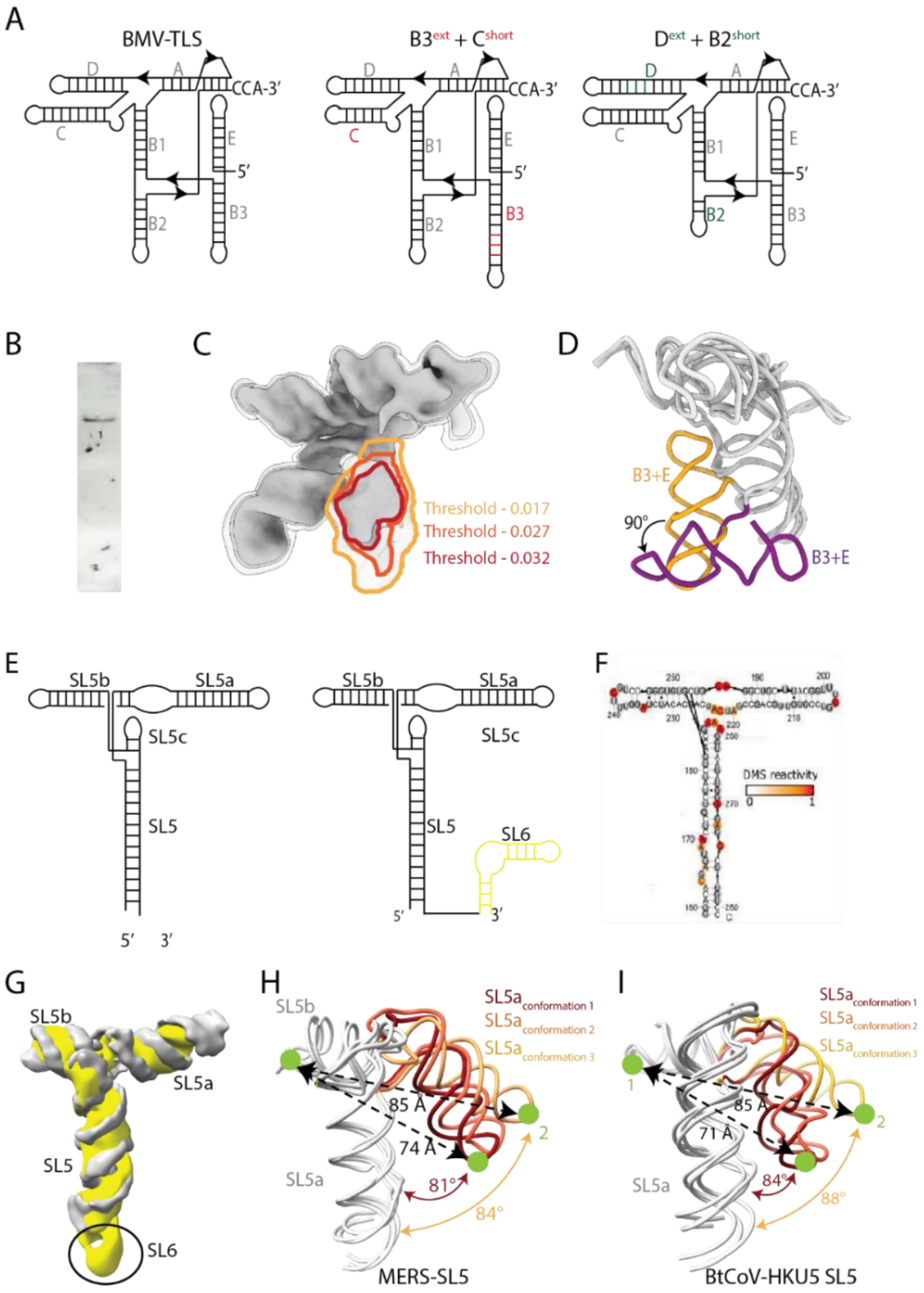

**Figure 4. Strategic construct design enabled modeling dynamic viral domains at medium resolution. (A)** Secondary structure map of BMV-TLS WT and two RNA helix engineered constructs $B3^{ext}$+$C^{short}$ and $D^{ext}$+$B2^{short}$. **(B)** Native PAGE gel image of WT BMV-TLS. **(C)** Cryo-EM maps of the WT BMV-TLS element wherein the B3+E RNA helix is highlighted with solid line. **(D)** Structure superposition of BMV-TLS TyrRS bound and unbound states showing 90º rotation of B3+E helix (orange and purple for unbound and bound protein structures respectively). **(E)** Secondary structure map of SARS-CoV2 SL5 (left) compared to the secondary structure map of SARS-CoV2 SL5 with the SL6 extension (right, in yellow). **(F)** Flexible nucleotides identified by chemical probing are indicated as red dots over the two secondary structure maps, supporting their modelling as single stranded nucleotides. **(G)** Superposition of the cryo-EM maps of SARS-CoV2 SL5 (grey) and SL5-6 (yellow). Density for SL6 is indicated by the black circle. **(H)** Superposition of three conformations of MERS SL5. Stem SL5a in conformation 1, 2, and 3 is displayed in maroon, tomato, and goldenrod, respectively. Its rotation is indicated by the solid lines (in °). The change in distance for hexaloop 1 are indicated by the dotted lines (in Å). **(I)** Superposition of three conformations of BtCoV-HKU5 SL5. Stem SL5a from conformation 1, 2, and 3 are displayed in maroon, tomato, and goldenrod, respectively. Its rotation is indicated by the solid lines (in °). The change in distance for hexaloop 1 are indicated by the dotted lines (in Å). Panel B was reprinted from Supplementary Figure 9 of Bonilla et al. (Bonilla et al., 2021), with permission from AAAS. It is not covered by the CC-BY 4.0 [https://creativecommons.org/licenses/by/4.0/] license and further reproduction of this panel would need permission from the copyright holder. Panel F is reproduced from Supplementary Figure 2 panel "SARS-CoV-2 SL5 domain Scarless M2-seq" from Kretsch, R. C., et al. (Kretsch et al., 2024).

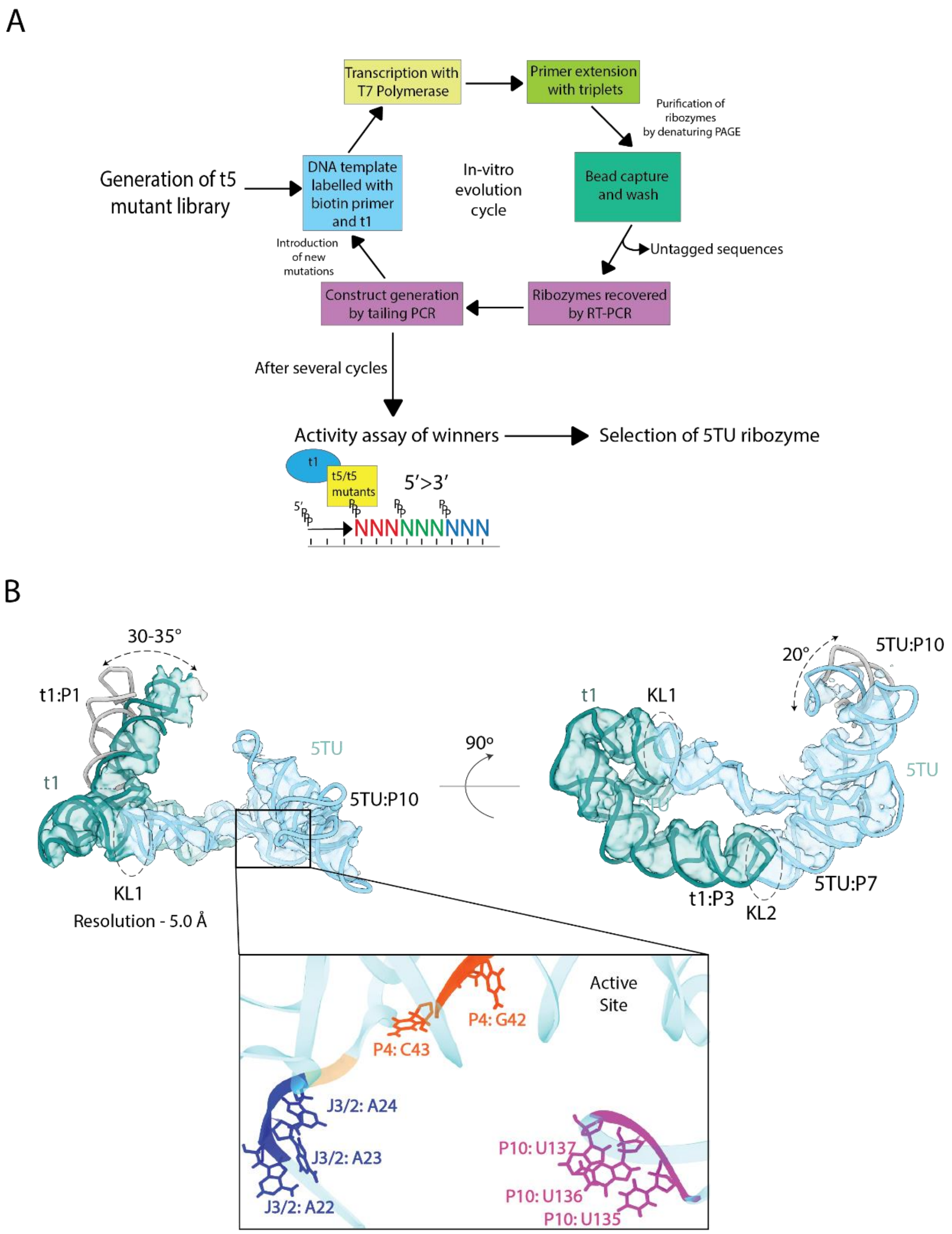

**Figure 5. *In vitro* evolution selected an efficient RNA polymerase ribozyme characterized by pronounced catalytically relevant dynamics. (A)** Schematic representation of the *in vitro* evolution cycle that was used to select the best performing RNA polymerase ribozyme, and specifically its 5TU catalytic subunit. **(B)** Cryo-EM map (EMD-40984) and structural coordinates (PDB id: 8T2P) of TPR polymerase subunits t1 (in teal) and 5TU (in light blue) in two orientations, rotated by 90° with respect to one another. Kissing loops (KL) 1 and 2, and helices P1 and P3 (from subunit t1), and P7 and P10 (from subunit 5TU) are indicated. The conformational changes of t1:P1 and 5TU:P10 helices are indicated by the dotted lines. The degree of motion is estimated from the figures reported in the original publication (McRae et al., 2024), because only one static set of coordinates is deposited in the PDB. The inset shows the active site of 5TU with all associated nucleotides in stick representation.

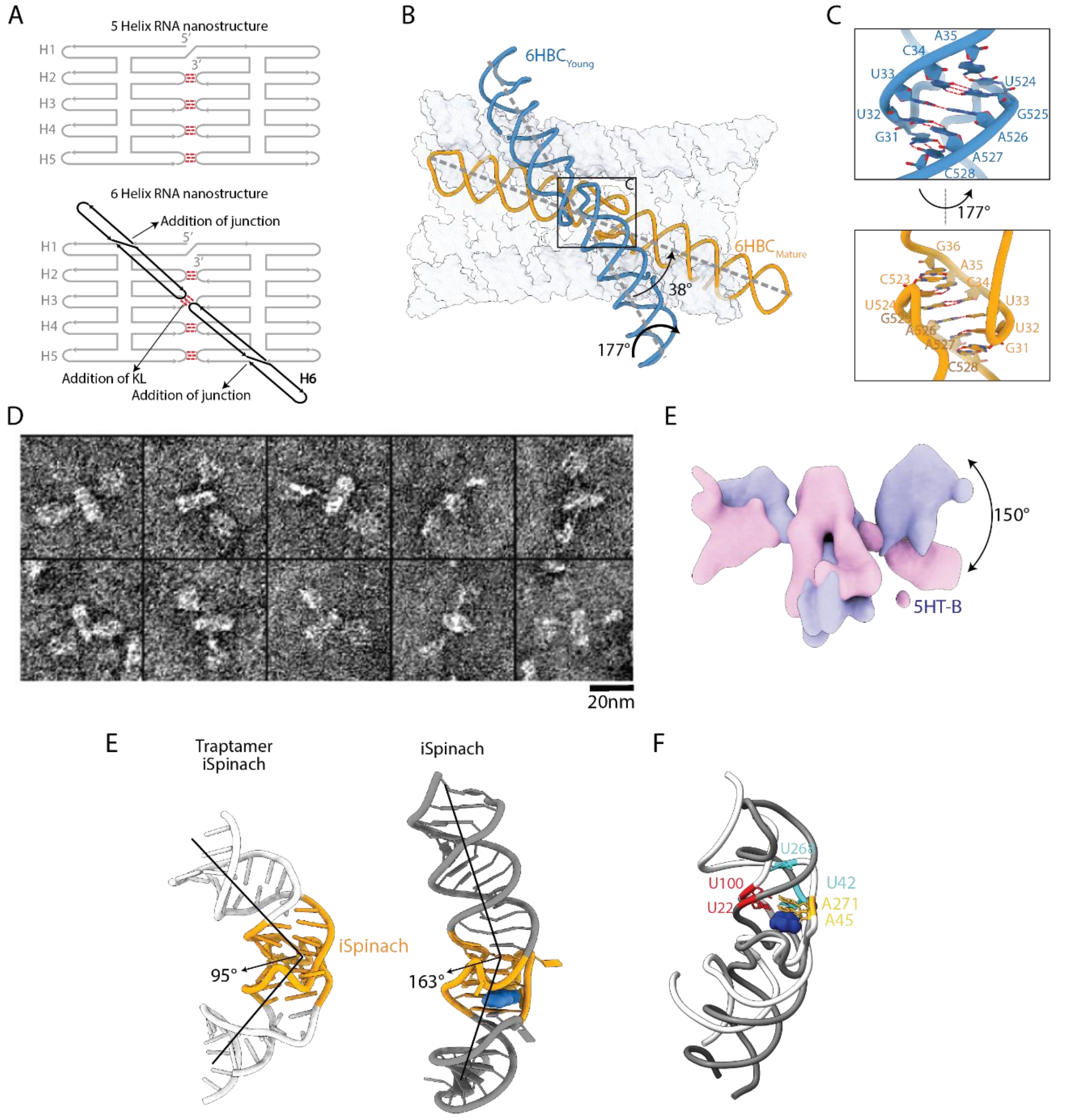

**Figure 6. Synthetic design of crossovers and fusions enabled the generation of dynamic RNA origami. (A)** Schematics depicting the engineering of kissing loops and junctions that enabled cryo-EM determination of the 6HBC origami. **(B)** Conformational rearrangement of H6 accompanying the transition of the 6HBC origami from the young to mature state. **(C)** Conformational rearrangement of the H6 kissing loop accompanying the transition of the 6HBC origami from the young (in blue) to the mature (in orange) state. A30 and A34 are shown in green; A522 and A526 are shown in orange. **(D)** Representative images of negatively stained 16HS-shaped particles. **(E)** Superposition of 16HS conformations 1 and 2. Flexibility of 5HT-B is also depicted with a curved arrow. **(F)** Comparison of the conformations of the iSpinach aptamer derived from the crystal structure (PDB: 5OB3, left) and the iSpinach aptamer model built from the cryo-EM density map (right). **(G)** Cartoon representations compare the iSpinach binding pocket between the crystal structure and cryo-EM structure. Relative to the crystal structure, the binding pocket of iSpinach in the Traptamer model is markedly contorted. Residues are colored as follows: U22 (crystal structure) and U100 (cryo-EM structure) in red; A45 (crystal structure) and A271 (cryo-EM structure) in orange; U42 (crystal structure) and U268 (cryo-EM structure) in cyan; DFHBI-1T is shown as blue spheres. Figure 6D was reprinted from Figure 5, panel B of (McRae et al., 2023) with permission from Springer Nature. It is not covered by the CC-BY 4.0 [https://creativecommons.org/licenses/by/4.0/] license and further reproduction of this panel would need permission from the copyright holder.

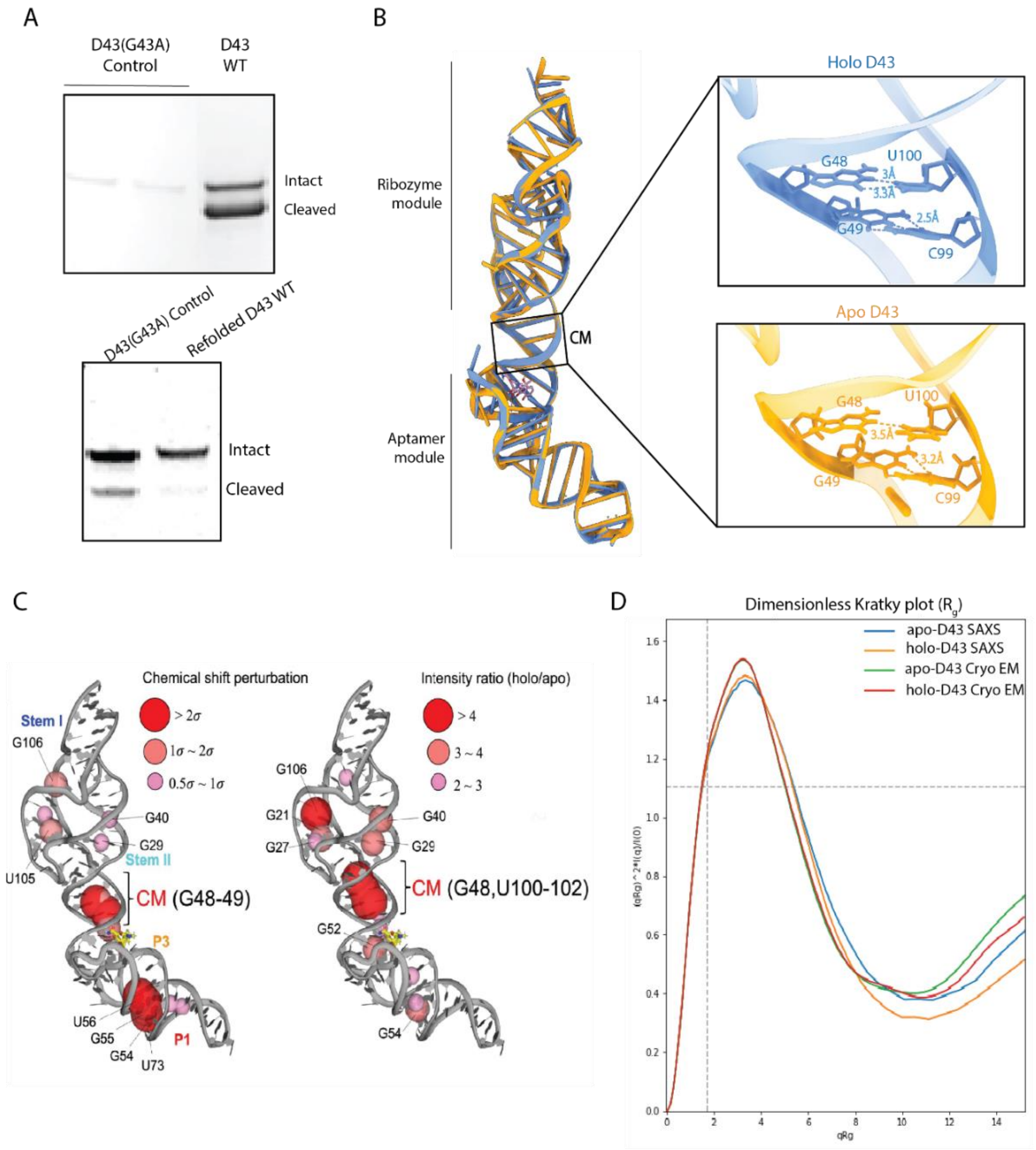

**Figure 7. Integration of cryo-EM with SAXS and NMR enabled the visualization of apo- and ligand-bound D43 aptazyme dynamics. (A)** Denaturing gel electrophoresis of purified D43 (left panel) and refolded D43 (right panel). **(B)** Superposition of the cryo-EM structures of holo-D43 (PDB ID: 8T5O; shown in blue) and apo-D43 (PDB ID: 8TKJ; shown in orange), which differ by an RMSD of 1.4 Å. The three modules of the D43 aptazyme are labelled, as the ribozyme module, aptamer module, and communication module (CM), respectively. The inset shows a zoomed view into the CM in holo-D43 and apo-D43, along with the distances between G48-U100 and G49-C99. The increase in the distance between G48-U100 and loss of planarity in G49-C99 lead to breakage of base pairing in apo-D43. **(C)** The location and extent of the NMR chemical shift perturbations (left panel) and intensity ratio differences (right panel) as indicated by the size and color of spheres. Large changes in both chemical shift and peak intensity are observed within the CM, most notably the G48•U100 wobble pair. **(D)** Kratky plot for apo-D43 and holo-D43 from SAXS/WAXS experiments and as back-calculated from the cryo-EM structures (PDB ids 8TKJ and 8T5O, respectively). The plot was generated using BioXTAS RAW 2.4.1 (Hopkins, 2024). Panels A and C are reproduced with permission from Supplementary Figure S1 panels B and D and from Figure 3 panel B, respectively, from Stagno, J. R., et al. (2025), Nucleic Acids Research, with permission from Nucleic Acids Research (Stagno et al., 2025). They are not covered by the CC-BY 4.0 [https://creativecommons.org/licenses/by/4.0/] license and further reproduction of this panel would need permission from the copyright holder."

# Tables

## Table 1. Key properties of the targets discussed in the case studies.

| Case study | RNA Specimen | RNA length (nt) | Molecular Weight (kDa) | RNA folding buffer | Purification method | Biochemical characterisation | Grid | Data processing | Resolution (Å) | Dynamics | Functional implications of dynamics |
|---|---|---|---|---|---|---|---|---|---|---|---|
| 1. | *Tetrahymena* group I intron | ~450 | 150 | 40 mM Tris–HCl, pH 7.0–7.9, 0.01% TritonX-100, 6–20 mM $MgCl_2$, 2 mM spermidine, 10 mM DTT | Non-denaturing purification | Splicing assay | 200 mesh Au R2/1 Quantifoil | EMAN2, Relion | 2.7-4.1 | Movement of P1 helix | To transfer to 5'-exon closer to active site for cleavage |
| 2. | *O. iheyensis* group II intron | ~300-400 | ~100 | 10 mM $MgCl_2$, 150 mM KCl, 5 mM Na-MES pH 6.5 | Non-denaturing purification | SAXS, SEC, MP, splicing assay | 300 mesh Cu 1.2/1.3 Quantifoil | *CryoSPARC*, JANAS | 3.7-7 | Movement of D1c helix | To open group II intron core for positioning of domain 5 |
| 3. | Cobalamin riboswitch | 210 | 70 | 50 mM MES, pH 6.0, 10 mM KCl, 1 mM $MgCl_2$ | Denaturing purification | N/A | 300 mesh Au R1.2/1.3 Quantifoil | Relion, CryoSPARC, SIMPLE3 | 2.9-5.3 | Dynamics of dimer and P6 domain | Dynamics is important to bind to cobalamin ligand |
| 4. | Brome mosaic virus tRNA-like structure RNA | 171 | 55 | 50 mM Na-MOPS, pH 7.0, 10 mM $MgCl_2$ | Non-denaturing purification | Aminoacylation assay | 400 mesh C-flat R1.2/1.3 EMS | Relion, CryoSPARC | 4.3 | Movement of B3 + E helix | The helix dynamics is critical to bind to transfer RNA |
| | | | | | | | | | | | synthetase |
| 5. | SARS-CoV-2 SL5 | 124 | 40 | 50 mM Na–HEPES, pH 8.0, 10 mM $MgCl_2$ | Denaturing purification | Chemical probing | 200 mesh Cu R2/1 and 300 mesh Cu R1.2/1.3 Quantifoil | *CryoSPARC* | 4.7 | N/A | N/A |
| | SARS-CoV-1 SL5 | 143 | 46 | 50 mM Na–HEPES, pH 8.0, 10 mM $MgCl_2$ | Denaturing purification | Chemical probing | 300 mesh Cu R1.2/1.3 Quantifoil | *CryoSPARC* | 7.1 | N/A | N/A |
| | MERS SL5 | 135 | 43 | 50 mM Na–HEPES, pH 8.0, 10 mM $MgCl_2$ | Denaturing purification | Chemical probing | 300 mesh Cu R1.2/1.3 Quantifoil | *CryoSPARC* | 6.4-6.9 | Dynamics of SL5a stem | N/A |
| | BtCoV-HKU5 SL5 | 135 | 43 | 50 mM Na–HEPES, pH 8.0, 10 mM $MgCl_2$ | Denaturing purification | Chemical probing | 300 mesh Cu R1.2/1.3 Quantifoil | *CryoSPARC* | 5.9-7.3 | Dynamics of SL5a stem | N/A |
| | HCoV-229E SL5 | 140 | 45 | 50 mM Na–HEPES, pH 8.0, 10 mM $MgCl_2$ | Denaturing purification | Chemical probing | 300 mesh Cu R1.2/1.3 Quantifoil | *CryoSPARC* | 6.5 | N/A | N/A |
| 6. | RNA polymerase ribozyme | 287 | 92 | 50 mM Tris–HCl, pH 8.0, 100 mM $MgCl_2$ | Denaturing purification | PAGE, negative staining | 300 mesh Au-flat 1.2/1.3 ProtoChips | *CryoSPARC* | 5.0 | Movement of P1 and P10 helices | To bind to the substrate to perform the replication |
| 7. | RNA origami nanostructure | 720 | 232 | 25 mM HEPES buffer | Non-denaturing | SAXS | 300 mesh Au-flat | *CryoSPARC*, Warp | 4.9-5.2 | Movement of H6 helix | Movement of the H6 helix is |
| | 6HBC | | | (pH 8.0), 50 mM KCl and 5 mM $MgCl_2$ | purification | | 1.2/1.3 ProtoChips | | | | critical for compaction into the native conformation of the nanostructure |
| | RNA origami nanostructure 16HS | 1.832 | ~600 | 25 mM HEPES buffer (pH 8.0), 50 mM KCl and 5 mM $MgCl_2$ | Non-denaturing purification | Negative staining | 300 mesh Au-flat 1.2/1.3 ProtoChips | IPET cryo-EM 3D reconstruction | 23-27 | Movement of the 5HT domains | N/A |
| | RNA origami nanostructure Traptamer | 374 | 121 | 40 mM Hepes pH 7.5, 50 mM KCl, and 5 mM $MgCl_2$ | Non-denaturing purification | Fluorescence measurement | 300 mesh Au-flat 1.2/1.3 ProtoChips | *CryoSPARC* | 5.4 | Bending of the iSpinach structure | To control the fluorescence from iSpinach |
| 8. | D43 aptazyme | 124 | ~40 | 10 mM Bis–Tris (pH 6.8), 100 mM KCl, and 1 mM $MgCl_2$ | Non-denaturing purification | SAXS and NMR | 300 mesh Au 1.2/1.3 Quantifoil | SIMPLE 3.0, *CryoSPARC*, Relion | 3.0-3.2 | Dynamics of the communication module | To transfer to ligand binding signal from aptamer to ribozyme module |

## References

Abramson, J., Adler, J., Dunger, J., Evans, R., Green, T., Pritzel, A., Ronneberger, O., Willmore, L., Ballard, A. J., Bambrick, J., Bodenstein, S. W., Evans, D. A., Hung, C. C., O'Neill, M., Reiman, D., Tunyasuvunakool, K., Wu, Z., Zemgulyte, A., Arvaniti, E.,…Jumper, J. M. (2024). Accurate structure prediction of biomolecular interactions with AlphaFold 3. *Nature*, *630*(8016), 493-500. https://doi.org/10.1038/s41586-024-07487-w

Aguilar, R., Spencer, K. B., Kesner, B., Rizvi, N. F., Badmalia, M. D., Mrozowich, T., Mortison, J. D., Rivera, C., Smith, G. F., Burchard, J., Dandliker, P. J., Patel, T. R., Nickbarg, E. B., & Lee, J. T. (2022). Targeting Xist with compounds that disrupt RNA structure and X inactivation. *Nature*, *604*(7904), 160-166. https://doi.org/10.1038/s41586-022-04537-z

Altman, S., Baer, M. F., Bartkiewicz, M., Gold, H., Guerrier-Takada, C., Kirsebom, L. A., Lumelsky, N., & Peck, K. (1989). Catalysis by the RNA subunit of RNase P--a minireview. *Gene*, *82*(1), 63-64. https://doi.org/10.1016/0378-1119(89)90030-9

Arney, J. W., Laederach, A., & Weeks, K. M. (2024). Visualizing RNA structure ensembles by single-molecule correlated chemical probing. *Curr Opin Struct Biol*, *88*, 102877. https://doi.org/10.1016/j.sbi.2024.102877

Arnold, E. B., Cohn, D., Bose, E., Klingler, D., Wolfe, G., & Jones, A. N. (2025). Investigating the interplay between RNA structural dynamics and RNA chemical probing experiments. *Nucleic Acids Res*, *53*(7). https://doi.org/10.1093/nar/gkaf290

Astrachan, L., & Volkin, E. (1958). Properties of ribonucleic acid turnover in T2-infected Escherichia coli. *Biochim Biophys Acta*, *29*(3), 536-544. https://doi.org/10.1016/0006-3002(58)90010-6

Attwater, J., Raguram, A., Morgunov, A. S., Gianni, E., & Holliger, P. (2018). Ribozyme-catalysed RNA synthesis using triplet building blocks. *Elife*, *7*. https://doi.org/10.7554/eLife.35255

Badepally, N. G., de Moura, T. R., Purta, E., Baulin, E. F., & Bujnicki, J. M. (2024). Cryo-EM Structure of raiA ncRNA From Clostridium Reveals a New RNA 3D Fold. *J Mol Biol*, *436*(23), 168833. https://doi.org/10.1016/j.jmb.2024.168833

Baek, M., McHugh, R., Anishchenko, I., Jiang, H., Baker, D., & DiMaio, F. (2024). Accurate prediction of protein-nucleic acid complexes using RoseTTAFoldNA. *Nat Methods*, *21*(1), 117-121. https://doi.org/10.1038/s41592-023-02086-5

Bernetti, M., & Bussi, G. (2023). Integrating experimental data with molecular simulations to investigate RNA structural dynamics. *Curr Opin Struct Biol*, *78*, 102503. https://doi.org/10.1016/j.sbi.2022.102503

Beusch, I., & Madhani, H. D. (2024). Understanding the dynamic design of the spliceosome. *Trends Biochem Sci*, *49*(7), 583-595. https://doi.org/10.1016/j.tibs.2024.03.012

Bhattacharjee, S., Abaeva, I. S., Brown, Z. P., Arhab, Y., Fallah, H., Hellen, C. U. T., Frank, J., & Pestova, T. V. (2026). The mechanism of ribosomal recruitment during translation initiation on the Type 2 encephalomyocarditis virus IRES. *EMBO J*, *45*(8), 2666-2693. https://doi.org/10.1038/s44318-026-00735-x

Biela, A., Hammermeister, A., Kaczmarczyk, I., Walczak, M., Koziej, L., Lin, T. Y., & Glatt, S. (2023). The diverse structural modes of tRNA binding and recognition. *J Biol Chem*, *299*(8), 104966. https://doi.org/10.1016/j.jbc.2023.104966

Boniecki, M. J., Lach, G., Dawson, W. K., Tomala, K., Lukasz, P., Soltysinski, T., Rother, K. M., & Bujnicki, J. M. (2016). SimRNA: a coarse-grained method for RNA folding simulations and 3D structure prediction. *Nucleic Acids Res*, *44*(7), e63. https://doi.org/10.1093/nar/gkv1479

Bonilla, S. L., & Jang, K. (2024). Challenges, advances, and opportunities in RNA structural biology by Cryo-EM. *Curr Opin Struct Biol*, *88*, 102894. https://doi.org/10.1016/j.sbi.2024.102894

Bonilla, S. L., & Kieft, J. S. (2022). The promise of cryo-EM to explore RNA structural dynamics. *J Mol Biol*, *434*(18), 167802. https://doi.org/10.1016/j.jmb.2022.167802

Bonilla, S. L., Sherlock, M. E., MacFadden, A., & Kieft, J. S. (2021). A viral RNA hijacks host machinery using dynamic conformational changes of a tRNA-like structure. *Science*, *374*(6570), 955-960. https://doi.org/10.1126/science.abe8526

Bonilla, S. L., Vicens, Q., & Kieft, J. S. (2022). Cryo-EM reveals an entangled kinetic trap in the folding of a catalytic RNA. *Sci Adv*, *8*(34), eabq4144. https://doi.org/10.1126/sciadv.abq4144

Breaker, R. R., Harris, K. A., Lyon, S. E., Wencker, F. D. R., & Fernando, C. M. (2023). Evidence that OLE RNA is a component of a major stress-responsive ribonucleoprotein particle in extremophilic bacteria. *Mol Microbiol*, *120*(3), 324-340. https://doi.org/10.1111/mmi.15129

Brenner, S., Jacob, F., & Meselson, M. (1961). An unstable intermediate carrying information from genes to ribosomes for protein synthesis. *Nature*, *190*, 576-581. https://doi.org/10.1038/190576a0

Bussi, G., Bonomi, M., Gkeka, P., Sattler, M., Al-Hashimi, H. M., Auffinger, P., Duca, M., Foricher, Y., Incarnato, D., Jones, A. N., Kirmizialtin, S., Krepl, M., Orozco, M., Palermo, G., Pasquali, S., Salmon, L., Schwalbe, H., Westhof, E., & Zacharias, M. (2024). RNA dynamics from experimental and computational approaches. *Structure*, *32*(9), 1281-1287. https://doi.org/10.1016/j.str.2024.07.019

Caesar, J., Reboul, C. F., Machello, C., Kiesewetter, S., Tang, M. L., Deme, J. C., Johnson, S., Elmlund, D., Lea, S. M., & Elmlund, H. (2020). SIMPLE 3.0. Stream single-particle cryo-EM analysis in real time. *J Struct Biol X*, *4*, 100040. https://doi.org/10.1016/j.yjsbx.2020.100040

Cai, X., Zhou, K., Alvarez-Cabrera, A. L., Si, Z., Wang, H., He, Y., Li, C., & Zhou, Z. H. (2024). Structural Heterogeneity of the Rabies Virus Virion. *Viruses*, *16*(9). https://doi.org/10.3390/v16091447

Cao, H., Wang, Y., Zhang, N., Xia, S., Tian, P., Lu, L., Du, J., & Du, Y. (2022). Progress of CRISPR-Cas13 Mediated Live-Cell RNA Imaging and Detection of RNA-Protein Interactions. *Front Cell Dev Biol*, *10*, 866820. https://doi.org/10.3389/fcell.2022.866820

Carbone, C. E., Loveland, A. B., Gamper, H. B., Jr., Hou, Y. M., Demo, G., & Korostelev, A. A. (2021). Time-resolved cryo-EM visualizes ribosomal translocation with EF-G and GTP. *Nat Commun*, *12*(1), 7236. https://doi.org/10.1038/s41467-021-27415-0

Cate, J. H., Gooding, A. R., Podell, E., Zhou, K., Golden, B. L., Kundrot, C. E., Cech, T. R., & Doudna, J. A. (1996). Crystal structure of a group I ribozyme domain: principles of RNA packing. *Science*, *273*(5282), 1678-1685. http://www.ncbi.nlm.nih.gov/pubmed/8781224

Cech, T. R. (1990). Self-splicing of group I introns [Review]. *Annu Rev Biochem*, *59*, 543-568. https://doi.org/10.1146/annurev.bi.59.070190.002551

Chapman, H. N., Fromme, P., Barty, A., White, T. A., Kirian, R. A., Aquila, A., Hunter, M. S., Schulz, J., DePonte, D. P., Weierstall, U., Doak, R. B., Maia, F. R., Martin, A. V., Schlichting, I., Lomb, L., Coppola, N., Shoeman, R. L., Epp, S. W., Hartmann, R.,…Spence, J. C. (2011). Femtosecond X-ray protein nanocrystallography. *Nature*, *470*(7332), 73-77. https://doi.org/10.1038/nature09750

Chen, L. L., & Kim, V. N. (2024). Small and long non-coding RNAs: Past, present, and future. *Cell*, *187*(23), 6451-6485. https://doi.org/10.1016/j.cell.2024.10.024

Chen, M., & Ludtke, S. J. (2021). Deep learning-based mixed-dimensional Gaussian mixture model for characterizing variability in cryo-EM. *Nat Methods*, *18*(8), 930-936. https://doi.org/10.1038/s41592-021-01220-5

Chen, M., Toader, B., & Lederman, R. (2023). Integrating Molecular Models Into CryoEM Heterogeneity Analysis Using Scalable High-resolution Deep Gaussian Mixture Models. *J Mol Biol*, *435*(9), 168014. https://doi.org/10.1016/j.jmb.2023.168014

Chen, X., Wang, L., Xie, J., Nowak, J. S., Luo, B., Zhang, C., Jia, G., Zou, J., Huang, D., Glatt, S., Yang, Y., & Su, Z. (2025). RNA sample optimization for cryo-EM analysis. *Nat Protoc*, *20*(5), 1114-1157. https://doi.org/10.1038/s41596-024-01072-1

Chen, Y., & Pollack, L. (2016). SAXS studies of RNA: structures, dynamics, and interactions with partners. *Wiley Interdiscip Rev RNA*, *7*(4), 512-526. https://doi.org/10.1002/wrna.1349

Cheng, C. Y., Kladwang, W., Yesselman, J. D., & Das, R. (2017). RNA structure inference through chemical mapping after accidental or intentional mutations. *Proc Natl Acad Sci U S A*, *114*(37), 9876-9881. https://doi.org/10.1073/pnas.1619897114

Chillon, I., Marcia, M., Legiewicz, M., Liu, F., Somarowthu, S., & Pyle, A. M. (2015). Native purification and analysis of long RNAs. *Methods Enzymol*, *558*, 3-37. https://doi.org/10.1016/bs.mie.2015.01.008

Cousin, F. J., Lynch, D. B., Chuat, V., Bourin, M. J. B., Casey, P. G., Dalmasso, M., Harris, H. M. B., McCann, A., & O'Toole, P. W. (2017). A long and abundant non-coding RNA in Lactobacillus salivarius. *Microb Genom*, *3*(9), e000126. https://doi.org/10.1099/mgen.0.000126

de Moura, T. R., Purta, E., Bernat, A., Martin-Cuevas, E. M., Kurkowska, M., Baulin, E. F., Mukherjee, S., Nowak, J., Biela, A. P., Rawski, M., Glatt, S., Moreno-Herrero, F., & Bujnicki, J. M. (2024). Conserved structures and dynamics in 5'-proximal regions of Betacoronavirus RNA genomes. *Nucleic Acids Res*, *52*(6), 3419-3432. https://doi.org/10.1093/nar/gkae144

Ding, J., Deme, J. C., Stagno, J. R., Yu, P., Lea, S. M., & Wang, Y. X. (2023). Capturing heterogeneous conformers of cobalamin riboswitch by cryo-EM. *Nucleic Acids Res*, *51*(18), 9952-9960. https://doi.org/10.1093/nar/gkad651

Ding, J., Lee, Y. T., Bhandari, Y., Schwieters, C. D., Fan, L., Yu, P., Tarosov, S. G., Stagno, J. R., Ma, B., Nussinov, R., Rein, A., Zhang, J., & Wang, Y. X. (2023). Visualizing RNA conformational and architectural heterogeneity in solution. *Nat Commun*, *14*(1), 714. https://doi.org/10.1038/s41467-023-36184-x

Dube, N., Ramelot, T. A., Benavides, T. L., Huang, Y. J., Moult, J., Kryshtafovych, A., & Montelione, G. T. (2026). Modeling Alternative Conformational States in CASP16. *Proteins*, *94*(1), 330-347. https://doi.org/10.1002/prot.70065

Fang, X., Wang, J., O'Carroll, I. P., Mitchell, M., Zuo, X., Wang, Y., Yu, P., Liu, Y., Rausch, J. W., Dyba, M. A., Kjems, J., Schwieters, C. D., Seifert, S., Winans, R. E., Watts, N. R., Stahl, S. J., Wingfield, P. T., Byrd, R. A., Le Grice, S. F.,…Wang, Y. X. (2013). An unusual topological structure of the HIV-1 Rev response element. *Cell*, *155*(3), 594-605. https://doi.org/10.1016/j.cell.2013.10.008

Frauenfelder, H., Sligar, S. G., & Wolynes, P. G. (1991). The energy landscapes and motions of proteins. *Science*, *254*(5038), 1598-1603. https://doi.org/10.1126/science.1749933

Ganser, L. R., Kelly, M. L., Herschlag, D., & Al-Hashimi, H. M. (2019). The roles of structural dynamics in the cellular functions of RNAs. *Nat Rev Mol Cell Biol*, *20*(8), 474-489. https://doi.org/10.1038/s41580-019-0136-0

Garg, P., Feng, X., De, S., & Frank, J. (2025). Passage of ribosomes through microsprayer increases functional activity - Implications for activity assays in time-resolved cryo-EM. *J Struct Biol*, *217*(3), 108232. https://doi.org/10.1016/j.jsb.2025.108232

Garst, A. D., Edwards, A. L., & Batey, R. T. (2011). Riboswitches: structures and mechanisms. *Cold Spring Harb Perspect Biol*, *3*(6). https://doi.org/10.1101/cshperspect.a003533

Geary, C., Grossi, G., McRae, E. K. S., Rothemund, P. W. K., & Andersen, E. S. (2021). RNA origami design tools enable cotranscriptional folding of kilobase-sized nanoscaffolds. *Nat Chem*, *13*(6), 549-558. https://doi.org/10.1038/s41557-021-00679-1

Gros, F., Hiatt, H., Gilbert, W., Kurland, C. G., Risebrough, R. W., & Watson, J. D. (1961). Unstable ribonucleic acid revealed by pulse labelling of Escherichia coli. *Nature*, *190*, 581-585. https://doi.org/10.1038/190581a0

Groves, D., Hepp, C., Kapanidis, A. N., & Robb, N. C. (2023). Single-molecule FRET for virology: 20 years of insight into protein structure and dynamics. *Q Rev Biophys*, *56*, e3. https://doi.org/10.1017/S0033583523000021

Haack, D. B., Rudolfs, B., Jin, S., Khitun, A., Weeks, K. M., & Toor, N. (2025). Scaffold-enabled high-resolution cryo-EM structure determination of RNA. *Nat Commun*, *16*(1), 880. https://doi.org/10.1038/s41467-024-55699-5

Hampton, C. M., Strauss, J. D., Ke, Z., Dillard, R. S., Hammonds, J. E., Alonas, E., Desai, T. M., Marin, M., Storms, R. E., Leon, F., Melikyan, G. B., Santangelo, P. J., Spearman, P. W., & Wright, E. R. (2017). Correlated fluorescence microscopy and cryo-electron tomography of virus-infected or transfected mammalian cells. *Nat Protoc*, *12*(1), 150-167. https://doi.org/10.1038/nprot.2016.168

Hartmann, A., Sreenivasa, K., Schenkel, M., Chamachi, N., Schake, P., Krainer, G., & Schlierf, M. (2023). An automated single-molecule FRET platform for high-content, multiwell plate screening of biomolecular conformations and dynamics. *Nat Commun*, *14*(1), 6511. https://doi.org/10.1038/s41467-023-42232-3

He, Y., Zhong, J., Yang, Y., Gunsalus, R. P., Zhou, Z. H., & Feigon, J. (2026). Cryo-EM structures reveal a conserved architecture for raiA noncoding RNA. *Nucleic Acids Res*, *54*(5). https://doi.org/10.1093/nar/gkag185

Henzler-Wildman, K., & Kern, D. (2007). Dynamic personalities of proteins. *Nature*, *450*(7172), 964-972. https://doi.org/10.1038/nature06522

Herron, L., Qiu, Y., Verma, A., Sreyas Adury, V. S., John, R., Lee, S., Mehdi, S., Sanwal, D., Schneekloth, J. S., & Tiwary, P. (2026). Ab initio prediction of RNA structure ensembles with RNAnneal. *bioRxiv*. https://doi.org/10.64898/2026.02.01.703098

Hershey, A. D. (1953). Nucleic acid economy in bacteria infected with bacteriophage T2. *J Gen Physiol*, *37*(1), 1-23. https://doi.org/10.1085/jgp.37.1.1

Hopkins, J. B. (2024). BioXTAS RAW 2: new developments for a free open-source program for small-angle scattering data reduction and analysis. *J Appl Crystallogr*, *57*(Pt 1), 194-208. https://doi.org/10.1107/S1600576723011019

Huang, Z., Lin, W., Wang, L., & Su, Z. (2026). Molecular insights into dynamic RNA quaternary assemblies. *RNA Biol*, *23*(1), 1-13. https://doi.org/10.1080/15476286.2026.2644525

Jacob, F., & Monod, J. (1961). Genetic regulatory mechanisms in the synthesis of proteins. *J Mol Biol*, *3*, 318-356. https://doi.org/10.1016/s0022-2836(61)80072-7

Jadhav, S., Maiorca, M., Manigrasso, J., Saha, S., Rakitch, A., Muscat, S., Mulvaney, T., De Vivo, M., Topf, M., & Marcia, M. (2025). Dynamic assembly of a large multidomain ribozyme visualized by cryo-electron microscopy. *Nat Commun*, *16*(1), 10195. https://doi.org/10.1038/s41467-025-65502-8

Jadhav, S., & Marcia, M. (2025). Cryo-specimen preparation and imaging of highly-structured and dynamic large non-coding RNAs. *BMC Methods*, *2*(1), 26. https://doi.org/10.1186/s44330-025-00045-4

Jia, X., Zhang, C., Luo, B., Frandsen, J. K., Watkins, A. M., Li, K., Zhang, M., Wei, X., Yang, Y., Henkin, T. M., & Su, Z. (2023). Cryo-EM-guided engineering of T-box-tRNA modules with enhanced selectivity and sensitivity in translational regulation. *bioRxiv*. https://doi.org/10.1101/2023.02.28.530422

Jones, C. P., & Ferre-D'Amare, A. R. (2025a). Crystallographic and cryoEM analyses reveal SARS-CoV-2 SL5 is a mobile T-shaped four-way junction with deep pockets. *Rna*, *31*(7), 949-960. https://doi.org/10.1261/rna.080413.125

Jones, C. P., & Ferre-D'Amare, A. R. (2025b). Structural switching dynamically controls the doubly pseudoknotted Rous sarcoma virus-programmed ribosomal frameshifting element. *Proc Natl Acad Sci U S A*, *122*(14), e2418418122. https://doi.org/10.1073/pnas.2418418122

Jones, C. P., & Ferre-D'Amare, A. R. (2026). Scaffolds with optimized quaternary symmetry for de novo cryoEM structure determination of small RNAs. *Nat Methods*. https://doi.org/10.1038/s41592-026-03016-x

Kappel, K., Zhang, K., Su, Z., Watkins, A. M., Kladwang, W., Li, S., Pintilie, G., Topkar, V. V., Rangan, R., Zheludev, I. N., Yesselman, J. D., Chiu, W., & Das, R. (2020). Accelerated cryo-EM-guided determination of three-dimensional RNA-only structures. *Nat Methods*, *17*(7), 699-707. https://doi.org/10.1038/s41592-020-0878-9

Kavita, K., & Breaker, R. R. (2023). Discovering riboswitches: the past and the future. *Trends Biochem Sci*, *48*(2), 119-141. https://doi.org/10.1016/j.tibs.2022.08.009
Khusainov, G., Standfuss, J., & Weinert, T. (2024). The time revolution in macromolecular crystallography. *Struct Dyn*, *11*(2), 020901. https://doi.org/10.1063/4.0000247
Kim, D. N., Thiel, B. C., Mrozowich, T., Hennelly, S. P., Hofacker, I. L., Patel, T. R., & Sanbonmatsu, K. Y. (2020). Zinc-finger protein CNBP alters the 3-D structure of lncRNA Braveheart in solution. *Nat Commun*, *11*(1), 148. https://doi.org/10.1038/s41467-019-13942-4
Kimanius, D., Forsberg, B. O., Scheres, S. H., & Lindahl, E. (2016). Accelerated cryo-EM structure determination with parallelisation using GPUs in RELION-2. *Elife*, *5*. https://doi.org/10.7554/eLife.18722
Kimanius, D., & Schwab, J. (2024). Confronting heterogeneity in cryogenic electron microscopy data: Innovative strategies and future perspectives with data-driven methods. *Curr Opin Struct Biol*, *86*, 102815. https://doi.org/10.1016/j.sbi.2024.102815
Kretsch, R. C., Albrecht, R., Andersen, E. S., Chen, H. A., Chiu, W., Das, R., Gezelle, J. G., Hartmann, M. D., Hobartner, C., Hu, Y., Jadhav, S., Johnson, P. E., Jones, C. P., Koirala, D., Kristoffersen, E. L., Largy, E., Lewicka, A., Mackereth, C. D., Marcia, M.,…Kryshtafovych, A. (2026). Functional Relevance of CASP16 Nucleic Acid Predictions as Evaluated by Structure Providers. *Proteins*, *94*(1), 51-78. https://doi.org/10.1002/prot.70043
Kretsch, R. C., Posani, E., Baulin, E. F., Bujnicki, J. M., Bussi, G., Cheatham, T. E., 3rd, Chen, S. J., Elofsson, A., Farsani, M. A., Fisher, O. N., Gromiha, M. M., Gupta, A., Hamada, M., Harini, K., Hu, G., Huang, D., Iwakiri, J., Jain, A., Kagaya, Y.,…Das, R. (2025). Blind prediction of complex water and ion ensembles around RNA in CASP16. *bioRxiv*. https://doi.org/10.1101/2025.11.03.685595
Kretsch, R. C., Wu, Y., Shabalina, S. A., Lee, H., Nye, G., Koonin, E. V., Gao, A., Chiu, W., & Das, R. (2025). Naturally ornate RNA-only complexes revealed by cryo-EM. *Nature*, *643*(8073), 1135-1142. https://doi.org/10.1038/s41586-025-09073-0
Kretsch, R. C., Xu, L., Zheludev, I. N., Zhou, X., Huang, R., Nye, G., Li, S., Zhang, K., Chiu, W., & Das, R. (2024). Tertiary folds of the SL5 RNA from the 5' proximal region of SARS-CoV-2 and related coronaviruses. *Proc Natl Acad Sci U S A*, *121*(10), e2320493121. https://doi.org/10.1073/pnas.2320493121
Kruger, K., Grabowski, P. J., Zaug, A. J., Sands, J., Gottschling, D. E., & Cech, T. R. (1982). Self-splicing RNA: autoexcision and autocyclization of the ribosomal RNA intervening sequence of Tetrahymena. *Cell*, *31*(1), 147-157. https://doi.org/10.1016/0092-8674(82)90414-7
Langeberg, C. J., & Kieft, J. S. (2023). A generalizable scaffold-based approach for structure determination of RNAs by cryo-EM. *Nucleic Acids Res*, *51*(20), e100. https://doi.org/10.1093/nar/gkad784
Languin-Cattoen, O., & Bussi, G. (2026). RNA Dynamics and Interactions Revealed Through Atomistic Simulations. *Annu Rev Phys Chem*. https://doi.org/10.1146/annurev-physchem-082624-013453
Levitz, T. S., Weckener, M., Fong, I., Naismith, J. H., Drennan, C. L., Brignole, E. J., Clare, D. K., & Darrow, M. C. (2022). Approaches to Using the Chameleon: Robust, Automated, Fast-Plunge cryoEM Specimen Preparation. *Front Mol Biosci*, *9*, 903148. https://doi.org/10.3389/fmolb.2022.903148
Li, J., Zhang, S., Zhang, D., & Chen, S. J. (2022). Vfold-Pipeline: a web server for RNA 3D structure prediction from sequences. *Bioinformatics*, *38*(16), 4042-4043. https://doi.org/10.1093/bioinformatics/btac426
Li, S., Palo, M. Z., Pintilie, G., Zhang, X., Su, Z., Kappel, K., Chiu, W., Zhang, K., & Das, R. (2022). Topological crossing in the misfolded Tetrahymena ribozyme resolved by cryo-EM. *Proc Natl Acad Sci U S A*, *119*(37), e2209146119. https://doi.org/10.1073/pnas.2209146119

Li, S., Palo, M. Z., Zhang, X., Pintilie, G., & Zhang, K. (2023). Snapshots of the second-step self-splicing of Tetrahymena ribozyme revealed by cryo-EM. *Nat Commun*, *14*(1), 1294. https://doi.org/10.1038/s41467-023-36724-5

Li, S., Su, Z., Lehmann, J., Stamatopoulou, V., Giarimoglou, N., Henderson, F. E., Fan, L., Pintilie, G. D., Zhang, K., Chen, M., Ludtke, S. J., Wang, Y. X., Stathopoulos, C., Chiu, W., & Zhang, J. (2019). Structural basis of amino acid surveillance by higher-order tRNA-mRNA interactions. *Nat Struct Mol Biol*, *26*(12), 1094-1105. https://doi.org/10.1038/s41594-019-0326-7

Li, Z., Zhu, J., Hong, X., Mu, J., Zheng, Z., Cui, T., Sun, Y., Wei, T., & Chen, H. F. (2025). DynaRNA: accurate dynamic RNA conformation ensemble generation with diffusion model. *Commun Biol*, *8*(1), 1472. https://doi.org/10.1038/s42003-025-08875-2

Ling, X., Golovenko, D., Gan, J., Ma, J., Korostelev, A. A., & Fang, W. (2025). Cryo-EM structure of a natural RNA nanocage. *Nature*, *644*(8078), 1107-1115. https://doi.org/10.1038/s41586-025-09262-x

Liu, D., Thelot, F. A., Piccirilli, J. A., Liao, M., & Yin, P. (2022). Sub-3-A cryo-EM structure of RNA enabled by engineered homomeric self-assembly. *Nat Methods*, *19*(5), 576-585. https://doi.org/10.1038/s41592-022-01455-w

Loveland, A. B., Demo, G., & Korostelev, A. A. (2020). Cryo-EM of elongating ribosome with EF-Tu*GTP elucidates tRNA proofreading. *Nature*, *584*(7822), 640-645. https://doi.org/10.1038/s41586-020-2447-x

Luo, B., Zhang, C., Ling, X., Mukherjee, S., Jia, G., Xie, J., Jia, X., Liu, L., Baulin, E. F., Luo, Y., Jiang, L., Dong, H., Wei, X., Bujnicki, J. M., & Su, Z. (2023). Cryo-EM reveals dynamics of Tetrahymena group I intron self-splicing. *Nature Catalysis*, *6*(4), 298-309. https://doi.org/10.1038/s41929-023-00934-3

Ma, H., Jia, X., Zhang, K., & Su, Z. (2022). Cryo-EM advances in RNA structure determination. *Signal Transduct Target Ther*, *7*(1), 58. https://doi.org/10.1038/s41392-022-00916-0

Maiorca, M., Jadhav, S., Sweeney, A., Redersborg, B., Marini, G., Mulvaney, T., Marcia, M., & Topf, M. (2026). Exploring Under-Resolved Functional Regions in Cryo-EM Datasets with JANAS. *Nat Struct Mol Biol*, *(in press)*. https://doi.org/10.21203/rs.3.rs-6535769/v1

Majumder, S., DeMott, C. M., Reverdatto, S., Burz, D. S., & Shekhtman, A. (2016). Total Cellular RNA Modulates Protein Activity. *Biochemistry*, *55*(32), 4568-4573. https://doi.org/10.1021/acs.biochem.6b00330

Marcia, M., & Pyle, A. M. (2012). Visualizing group II intron catalysis through the stages of splicing. *Cell*, *151*(3), 497-507. https://doi.org/10.1016/j.cell.2012.09.033

Marcia, M., & Pyle, A. M. (2014). Principles of ion recognition in RNA: insights from the group II intron structures. *RNA*, *20*(4), 516-527. https://doi.org/10.1261/rna.043414.113

McRae, E. K. S., Rasmussen, H. O., Liu, J., Boggild, A., Nguyen, M. T. A., Sampedro Vallina, N., Boesen, T., Pedersen, J. S., Ren, G., Geary, C., & Andersen, E. S. (2023). Structure, folding and flexibility of co-transcriptional RNA origami. *Nat Nanotechnol*, *18*(7), 808-817. https://doi.org/10.1038/s41565-023-01321-6

McRae, E. K. S., Wan, C. J. K., Kristoffersen, E. L., Hansen, K., Gianni, E., Gallego, I., Curran, J. F., Attwater, J., Holliger, P., & Andersen, E. S. (2024). Cryo-EM structure and functional landscape of an RNA polymerase ribozyme. *Proc Natl Acad Sci U S A*, *121*(3), e2313332121. https://doi.org/10.1073/pnas.2313332121

Michel, F., & Ferat, J. L. (1995). Structure and activities of group II introns. *Annu Rev Biochem*, *64*, 435-461. https://doi.org/10.1146/annurev.bi.64.070195.002251

Miescher, F. (1871). Ueber die chemische Zusammensetzung der Eiterzellen. *Hoppe-Seylers medizinisch-chemische Untersuchungen*(4), 441-460.

Monod, J., Pappenheimer, A. M., Jr., & Cohen-Bazire, G. (1952). [The kinetics of the biosynthesis of beta-galactosidase in Escherichia coli as a function of growth]. *Biochim Biophys Acta*, *9*(6), 648-660. https://doi.org/10.1016/0006-3002(52)90227-8

(La cinetique de la biosynthese de la beta-galactosidase chez E. coli consideree comme fonction de la croissance.)
Moser, F., Prazak, V., Mordhorst, V., Andrade, D. M., Baker, L. A., Hagen, C., Grunewald, K., & Kaufmann, R. (2019). Cryo-SOFI enabling low-dose super-resolution correlative light and electron cryo-microscopy. *Proc Natl Acad Sci U S A*, *116*(11), 4804-4809. https://doi.org/10.1073/pnas.1810690116
Nalewaj, M., & Szabat, M. (2022). Examples of Structural Motifs in Viral Genomes and Approaches for RNA Structure Characterization. *Int J Mol Sci*, *23*(24). https://doi.org/10.3390/ijms232415917
Nemeth, K., Bayraktar, R., Ferracin, M., & Calin, G. A. (2024). Non-coding RNAs in disease: from mechanisms to therapeutics. *Nat Rev Genet*, *25*(3), 211-232. https://doi.org/10.1038/s41576-023-00662-1
Nilsen, T. W. (2013). Gel purification of RNA. *Cold Spring Harb Protoc*, *2013*(2), 180-183. https://doi.org/10.1101/pdb.prot072942
Pardee, A. B. (1954). Nucleic Acid Precursors and Protein Synthesis. *Proc Natl Acad Sci U S A*, *40*(5), 263-270. https://doi.org/10.1073/pnas.40.5.263
Park, H. Y., Lim, H., Yoon, Y. J., Follenzi, A., Nwokafor, C., Lopez-Jones, M., Meng, X., & Singer, R. H. (2014). Visualization of dynamics of single endogenous mRNA labeled in live mouse. *Science*, *343*(6169), 422-424. https://doi.org/10.1126/science.1239200
Patt, E., Classen, S., Hammel, M., & Schneidman-Duhovny, D. (2025). Predicting RNA structure and dynamics with deep learning and solution scattering. *Biophys J*, *124*(3), 549-564. https://doi.org/10.1016/j.bpj.2024.12.024
Pokorna, P., Aupic, J., Fica, S. M., & Magistrato, A. (2025). Decoding Spliceosome Dynamics through Computation and Experiment. *Chem Rev*, *125*(20), 9807-9833. https://doi.org/10.1021/acs.chemrev.5c00374
Pollack, L. (2011). SAXS studies of ion-nucleic acid interactions. *Annu Rev Biophys*, *40*, 225-242. https://doi.org/10.1146/annurev-biophys-042910-155349
Puerta-Fernandez, E., Barrick, J. E., Roth, A., & Breaker, R. R. (2006). Identification of a large noncoding RNA in extremophilic eubacteria. *Proc Natl Acad Sci U S A*, *103*(51), 19490-19495. https://doi.org/10.1073/pnas.0607493103
Punjani, A., & Fleet, D. J. (2021). 3D variability analysis: Resolving continuous flexibility and discrete heterogeneity from single particle cryo-EM. *J Struct Biol*, *213*(2), 107702. https://doi.org/10.1016/j.jsb.2021.107702
Punjani, A., Rubinstein, J. L., Fleet, D. J., & Brubaker, M. A. (2017). cryoSPARC: algorithms for rapid unsupervised cryo-EM structure determination. *Nat Methods*, *14*(3), 290-296. https://doi.org/10.1038/nmeth.4169
Qin, B., Lauer, S. M., Balke, A., Vieira-Vieira, C. H., Burger, J., Mielke, T., Selbach, M., Scheerer, P., Spahn, C. M. T., & Nikolay, R. (2023). Cryo-EM captures early ribosome assembly in action. *Nat Commun*, *14*(1), 898. https://doi.org/10.1038/s41467-023-36607-9
Rich, A., & Watson, J. D. (1954). Physical studies on ribonucleic acid. *Nature*, *173*(4412), 995-996. https://doi.org/10.1038/173995a0
Rickgauer, J. P., Choi, H., Moore, A. S., Denk, W., & Lippincott-Schwartz, J. (2024). Structural dynamics of human ribosomes in situ reconstructed by exhaustive high-resolution template matching. *Mol Cell*, *84*(24), 4912-4928 e4917. https://doi.org/10.1016/j.molcel.2024.11.003
Robertus, J. D., Ladner, J. E., Finch, J. T., Rhodes, D., Brown, R. S., Clark, B. F., & Klug, A. (1974). Structure of yeast phenylalanine tRNA at 3 A resolution. *Nature*, *250*(467), 546-551. https://doi.org/10.1038/250546a0
Rohou, A., & Grigorieff, N. (2015). CTFFIND4: Fast and accurate defocus estimation from electron micrographs. *J Struct Biol*, *192*(2), 216-221. https://doi.org/10.1016/j.jsb.2015.08.008
Russo, C. J., & Passmore, L. A. (2014). Electron microscopy: Ultrastable gold substrates for electron cryomicroscopy. *Science*, *346*(6215), 1377-1380. https://doi.org/10.1126/science.1259530

Sampedro Vallina, N., McRae, E. K. S., Geary, C., & Andersen, E. S. (2023). An RNA Paranemic Crossover Triangle as A 3D Module for Cotranscriptional Nanoassembly. *Small*, *19*(13), e2204651. https://doi.org/10.1002/smll.202204651

Scheres, S. H. (2012). RELION: implementation of a Bayesian approach to cryo-EM structure determination. *J Struct Biol*, *180*(3), 519-530. https://doi.org/10.1016/j.jsb.2012.09.006

Smith, A. M., Li, Y., Velarde, A., Cheng, Y., & Frankel, A. D. (2024). The HIV-1 Nuclear Export Complex Reveals the Role of RNA in Crm1 Cargo Recognition. *bioRxiv*. https://doi.org/10.1101/2024.09.22.614349

Spitale, R. C., & Incarnato, D. (2023). Probing the dynamic RNA structurome and its functions. *Nature reviews Genetics*, *24*(3), 178-196. https://doi.org/10.1038/s41576-022-00546-w

Stagno, J. R., Deme, J. C., Dwivedi, V., Lee, Y. T., Lee, H. K., Yu, P., Chen, S. Y., Fan, L., Degenhardt, M. F. S., Chari, R., Young, H. A., Lea, S. M., & Wang, Y. X. (2025). Structural investigation of an RNA device that regulates PD-1 expression in mammalian cells. *Nucleic Acids Res*, *53*(5). https://doi.org/10.1093/nar/gkaf156

Stagno, J. R., Liu, Y., Bhandari, Y. R., Conrad, C. E., Panja, S., Swain, M., Fan, L., Nelson, G., Li, C., Wendel, D. R., White, T. A., Coe, J. D., Wiedorn, M. O., Knoska, J., Oberthuer, D., Tuckey, R. A., Yu, P., Dyba, M., Tarasov, S. G.,…Wang, Y. X. (2017). Structures of riboswitch RNA reaction states by mix-and-inject XFEL serial crystallography. *Nature*, *541*(7636), 242-246. https://doi.org/10.1038/nature20599

Tang, G., Peng, L., Baldwin, P. R., Mann, D. S., Jiang, W., Rees, I., & Ludtke, S. J. (2007). EMAN2: an extensible image processing suite for electron microscopy. *J Struct Biol*, *157*(1), 38-46. https://doi.org/10.1016/j.jsb.2006.05.009

Thess, A., Hoerr, I., Panah, B. Y., Jung, G., & Dahm, R. (2021). Historic nucleic acids isolated by Friedrich Miescher contain RNA besides DNA. *Biol Chem*, *402*(10), 1179-1185. https://doi.org/10.1515/hsz-2021-0226

Toor, N., Rajashankar, K., Keating, K. S., & Pyle, A. M. (2008). Structural basis for exon recognition by a group II intron. *Nat Struct Mol Biol*, *15*(11), 1221-1222. https://doi.org/10.1038/nsmb.1509

Torino, S., Dhurandhar, M., Stroobants, A., Claessens, R., & Efremov, R. G. (2023). Time-resolved cryo-EM using a combination of droplet microfluidics with on-demand jetting. *Nat Methods*, *20*(9), 1400-1408. https://doi.org/10.1038/s41592-023-01967-z

Uroda, T., Anastasakou, E., Rossi, A., Teulon, J. M., Pellequer, J. L., Annibale, P., Pessey, O., Inga, A., Chillon, I., & Marcia, M. (2019). Conserved Pseudoknots in lncRNA MEG3 Are Essential for Stimulation of the p53 Pathway. *Mol Cell*, *75*(5), 982-995 e989. https://doi.org/10.1016/j.molcel.2019.07.025

Uroda, T., Chillon, I., Annibale, P., Teulon, J. M., Pessey, O., Karuppasamy, M., Pellequer, J. L., & Marcia, M. (2020). Visualizing the functional 3D shape and topography of long noncoding RNAs by single-particle atomic force microscopy and in-solution hydrodynamic techniques. *Nat Protoc*, *15*(6), 2107-2139. https://doi.org/10.1038/s41596-020-0323-7

Vallina, N. S., McRae, E. K. S., Geary, C., & Andersen, E. S. (2024). An RNA origami robot that traps and releases a fluorescent aptamer. *Sci Adv*, *10*(12), eadk1250. https://doi.org/10.1126/sciadv.adk1250

Vicens, Q., & Kieft, J. S. (2022). Thoughts on how to think (and talk) about RNA structure. *Proc Natl Acad Sci U S A*, *119*(17), e2112677119. https://doi.org/10.1073/pnas.2112677119

Waldsich, C., & Pyle, A. M. (2008). A kinetic intermediate that regulates proper folding of a group II intron RNA. *J Mol Biol*, *375*(2), 572-580. https://doi.org/10.1016/j.jmb.2007.10.052

Wallace, J. G., Zhou, Z., & Breaker, R. R. (2012). OLE RNA protects extremophilic bacteria from alcohol toxicity. *Nucleic Acids Res*, *40*(14), 6898-6907. https://doi.org/10.1093/nar/gks352

Wang, L., Xie, J., Gong, T., Wu, H., Tu, Y., Peng, X., Shang, S., Jia, X., Ma, H., Zou, J., Xu, S., Zheng, X., Zhang, D., Liu, Y., Zhang, C., Luo, Y., Huang, Z., Shao, B., Ying, B.,…Su, Z. (2025). Cryo-EM reveals mechanisms of natural RNA multivalency. *Science*, *388*(6746), 545-550. https://doi.org/10.1126/science.adv3451

Wang, L., Xie, J., Zhang, C., Zou, J., Huang, Z., Shang, S., Chen, X., Yang, Y., Liu, J., Dong, H., Huang, D., & Su, Z. (2025). Structural basis of circularly permuted group II intron self-splicing. *Nat Struct Mol Biol*. https://doi.org/10.1038/s41594-025-01484-x

Weber, M., Erichson, F., Antczak, M., Schumann, V., Meitzner, J., Zok, T., Steffen, F. D., Szachniuk, M., & Borner, R. (2026). FRET-guided selection of RNA 3D structures. *Nucleic Acids Res*, *54*(5). https://doi.org/10.1093/nar/gkag147

Wlodarski, T., Streit, J. O., Mitropoulou, A., Cabrita, L. D., Vendruscolo, M., & Christodoulou, J. (2024). Bayesian reweighting of biomolecular structural ensembles using heterogeneous cryo-EM maps with the cryoENsemble method. *Sci Rep*, *14*(1), 18149. https://doi.org/10.1038/s41598-024-68468-7

Xia, X., Sung, P. Y., Martynowycz, M. W., Gonen, T., Roy, P., & Zhou, Z. H. (2024). RNA genome packaging and capsid assembly of bluetongue virus visualized in host cells. *Cell*, *187*(9), 2236-2249 e2217. https://doi.org/10.1016/j.cell.2024.03.007

Xu, L., Liu, T., Chung, K., & Pyle, A. M. (2023). Structural insights into intron catalysis and dynamics during splicing. *Nature*, *624*(7992), 682-688. https://doi.org/10.1038/s41586-023-06746-6

Yang, W., Yi, R., Yao, J., Gao, Y., Li, S., Gong, Q., & Zhang, K. (2025). Structural insights into dynamics of the BMV TLS aminoacylation. *Nat Commun*, *16*(1), 1276. https://doi.org/10.1038/s41467-025-56612-4

Yang, Y., Liu, S., Egloff, S., Eichhorn, C. D., Hadjian, T., Zhen, J., Kiss, T., Zhou, Z. H., & Feigon, J. (2022). Structural basis of RNA conformational switching in the transcriptional regulator 7SK RNP. *Mol Cell*, *82*(9), 1724-1736 e1727. https://doi.org/10.1016/j.molcel.2022.03.001

Ycas, M., & Vincent, W. S. (1960). A Ribonucleic Acid Fraction from Yeast Related in Composition to Desoxyribonucleic Acid. *Proc Natl Acad Sci U S A*, *46*(6), 804-811. https://doi.org/10.1073/pnas.46.6.804

Yesselman, J. D., Eiler, D., Carlson, E. D., Gotrik, M. R., d'Aquino, A. E., Ooms, A. N., Kladwang, W., Carlson, P. D., Shi, X., Costantino, D. A., Herschlag, D., Lucks, J. B., Jewett, M. C., Kieft, J. S., & Das, R. (2019). Computational design of three-dimensional RNA structure and function. *Nat Nanotechnol*, *14*(9), 866-873. https://doi.org/10.1038/s41565-019-0517-8

Zhang, K., Keane, S. C., Su, Z., Irobalieva, R. N., Chen, M., Van, V., Sciandra, C. A., Marchant, J., Heng, X., Schmid, M. F., Case, D. A., Ludtke, S. J., Summers, M. F., & Chiu, W. (2018). Structure of the 30 kDa HIV-1 RNA Dimerization Signal by a Hybrid Cryo-EM, NMR, and Molecular Dynamics Approach. *Structure*, *26*(3), 490-498 e493. https://doi.org/10.1016/j.str.2018.01.001

Zhang, K., Li, S., Kappel, K., Pintilie, G., Su, Z., Mou, T. C., Schmid, M. F., Das, R., & Chiu, W. (2019). Cryo-EM structure of a 40 kDa SAM-IV riboswitch RNA at 3.7 A resolution. *Nat Commun*, *10*(1), 5511. https://doi.org/10.1038/s41467-019-13494-7

Zhang, K., Zheludev, I. N., Hagey, R. J., Wu, M. T., Haslecker, R., Hou, Y. J., Kretsch, R., Pintilie, G. D., Rangan, R., Kladwang, W., Li, S., Pham, E. A., Bernardin-Souibgui, C., Baric, R. S., Sheahan, T. P., V, D. S., Glenn, J. S., Chiu, W., & Das, R. (2020). Cryo-electron Microscopy and Exploratory Antisense Targeting of the 28-kDa Frameshift Stimulation Element from the SARS-CoV-2 RNA Genome. *bioRxiv*. https://doi.org/10.1101/2020.07.18.209270

Zhang, S., Yi, R., An, L., Liu, J., Yao, X., Li, S., & Zhang, K. (2025). Structural insights into higher-order natural RNA-only multimers. *Nat Struct Mol Biol*, *32*(10), 2012-2021. https://doi.org/10.1038/s41594-025-01650-1

Zhang, X., Li, S., Pintilie, G., Palo, M. Z., & Zhang, K. (2023). Snapshots of the first-step self-splicing of Tetrahymena ribozyme revealed by cryo-EM. *Nucleic Acids Res*, *51*(3), 1317-1325. https://doi.org/10.1093/nar/gkac1268

Zhang, X., Li, S., & Zhang, K. (2024). Cryo-EM: A window into the dynamic world of RNA molecules. *Curr Opin Struct Biol*, *88*, 102916. https://doi.org/10.1016/j.sbi.2024.102916

Zheng, S. Q., Palovcak, E., Armache, J. P., Verba, K. A., Cheng, Y., & Agard, D. A. (2017). MotionCor2: anisotropic correction of beam-induced motion for improved cryo-electron microscopy. *Nat Methods*, *14*(4), 331-332. https://doi.org/10.1038/nmeth.4193

Zhong, E. D., Bepler, T., Berger, B., & Davis, J. H. (2021). CryoDRGN: reconstruction of heterogeneous cryo-EM structures using neural networks. *Nat Methods*, *18*(2), 176-185. https://doi.org/10.1038/s41592-020-01049-4